%% file: main.tex
\documentclass[pre,floatfix,longbibliography,superscriptaddress,onecolumn,final,notitlepage]{revtex4-1}

\usepackage[utf8]{inputenc} 
\usepackage[T1]{fontenc}
\usepackage{multirow} 
\usepackage{subcaption} 
\usepackage{tabularx}
\usepackage{graphicx,caption} 
\usepackage{amsfonts,amsmath,amssymb}
\usepackage{comment} 
\usepackage{lineno}
\usepackage{hyperref} 
\usepackage[explicit]{titlesec}
\usepackage{cleveref} 

\input{preamble_math.tex}

\usepackage{algorithm,algorithmic} 
\usepackage[algoruled,algo2e]{algorithm2e}
\usepackage{sidecap}
\usepackage{boldline}
\usepackage{setspace}
\usepackage{xcolor}
\usepackage{textcomp}
\usepackage{marvosym}  

\usepackage{tabu}  
\usepackage{transparent}

\newcommand{\dmacd}{\mbox{{\small DynACD}}}
\newcommand{\dmcd}{\mbox{{\small DynCD}}}

\newcommand{\acd}{\mbox{{\small ACD}}}  
\newcommand{\eum}{\mbox{{\small EU-monthly}}}

\usepackage[normalem]{ulem} % to use nix

\usepackage{changepage}

\usepackage{tabu}

\usepackage{cleveref}
\crefname{equation}{Eq.}{Eqs.} 
\crefname{section}{Sec.}{Secs.}
\crefname{figure}{Fig.}{Figs.}

\begin{document} 
\title{Community detection and  anomaly prediction  in dynamic networks}

\author{Hadiseh Safdari}
	\email{hadiseh.safdari@tuebingen.mpg.de} 
	\affiliation{Max Planck Institute for Intelligent Systems, Cyber Valley, Tuebingen 72076, Germany}

\author{Caterina De Bacco}
	\email{caterina.debacco@tuebingen.mpg.de} 
	\affiliation{Max Planck Institute for Intelligent Systems, Cyber Valley, Tuebingen 72076, Germany}

\maketitle

\section*{Abstract}    

Anomaly detection is an essential task in the analysis of dynamic networks, offering early warnings of abnormal behavior.  We present a principled approach to detect anomalies in dynamic networks that integrates community structure as a foundational model for regular behavior. Our model identifies anomalies as irregular edges while capturing structural changes. Our approach leverages a Markovian framework for temporal transitions and latent variables for community and anomaly detection, inferring hidden parameters to detect unusual interactions. Evaluations on synthetic and real-world datasets show strong anomaly detection across various scenarios. In a case study on professional football player transfers, we detect patterns influenced by club wealth and country, as well as unexpected transactions both within and across community boundaries.  This work provides a framework for adaptable anomaly detection, highlighting the value of integrating domain knowledge with data-driven techniques for improved interpretability and robustness in complex networks.

\section{Introduction}   
Dynamic networks are ubiquitous in real-world applications, ranging from social networks to biological systems and transportation networks \citep{Michail2018,Skarding2021}. As these networks evolve over time, detecting anomalies in the networks becomes increasingly important for identifying critical events, predicting failures, and uncovering hidden patterns \citep{Ma2023Akoglu,Miao2023,Ranshous2015survey}. Anomaly detection in dynamic networks involves identifying objects (nodes, edges, subgraphs) that deviate significantly from the expected behavior of the network, given its past behavior. This process  enables the detection of irregularities that may be indicative of important changes or abnormalities within the network structure or dynamics. For instance, network intrusions  can result in confidential personal information being leaked to hackers. By utilising anomaly detection techniques, one can acquire a more profound understanding of the behavior and evolution of dynamic networks, allowing informed decision-making and necessary courses of action  \citep{OliveiraPinto2022,Villegas-Ch2023}.

Current approaches to detecting anomalies in dynamic graphs include machine-learning techniques, and graph-based algorithms \citep{YuNetWalk2018,Ma2023Akoglu,Wang2019,Zheng2019OCAN,Zheng2019AddGraph}.  
Although deep learning methods show potential in handling the complexity of graph data, they are not without limitations  \citep{Ma2023Akoglu}.  
A significant obstacle is the models' lack of interpretability, which leads to challenges in understanding the reasoning behind anomaly detection decisions, an essential aspect in applications that require explanations.

Moreover, these models heavily rely on large labeled datasets for effective training, which becomes impractical in dynamic network environments due to the laborious and expensive process of data collection and labeling.  Furthermore, the high computational complexity of deep learning models poses a barrier to real-time anomaly detection in dynamic networks, where timely identification is of utmost importance. In addition, 
many dynamic graph anomaly detection techniques in deep learning rely on models like DeepWalk \citep{PerozziDeepWalk2014}, and GCN \citep{kipf2017semisupervised} that are primarily designed for static graphs. As a result, they do not effectively utilize evolving patterns in attributes for the detection task \citep{Ma2023Akoglu}.

In addition to deep learning techniques, graph-based anomaly detection techniques, where nodes and edges represent objects and their relationships,  have also been explored for anomaly edge detection in dynamic networks \citep{Ma2023Akoglu}. However, previous works in this area, which rely on statistical models, often face limitations in accurately identifying unknown anomalies and struggle with scalability issues in large networks. As the complexity and scale of dynamic networks continue to grow, there is an increasing need for principled  approaches that can improve both the accuracy and efficiency of anomaly detection.
Furthermore, the models in this category do not explicitly model anomalies. Instead of systematically incorporating anomalies within their framework, these models treat anomalies as irregularities that occur outside of the standard data patterns  \citep{SedanSpot2018,Li2009}.

In this paper, we introduce a principled approach that combines community detection and anomaly prediction,  
to learn patterns of regular behavior in the network structure and identify deviations, i.e.,  anomalous behavior. 
 In many complex networks, nodes are clustered into communities of similar nodes \citep{HOLLAND1983109,Ball2011,Kirkley2022}. For example, in transportation networks, communities might be clusters of closely connected airports, showing busy routes or regional hubs. Similarly, in online social networks, communities might be groups of users with shared interests or frequent interactions. These communities reveal how the network is organized and how its members interact. They help us understand the structure and functions of the underlying networks. Here, we exploit these insights to encode regular behavior in a probabilistic generative model for networks by means of community structure that evolve over time. By encoding deviations from this behavior as anomalies, we are able to estimate the probability that a certain edge is anomalous while also learning the hidden community structure.

The incorporation of community membership in this setting  presents  advantages over traditional anomaly detection methods  \citep{Ma2023Akoglu,YuNetWalk2018,SafdariJBD2022}. For instance, by focusing on specific communities within the network, our approach can provide more targeted and accurate anomaly detection, as anomalies often manifest within localized regions rather than affecting the entire network. On the other hand, an evolving community structure can explain the underlying dynamics of the network effectively \citep{peixoto2019modelling,peixoto2017modelling,matias2015statistical}, enabling us to detect anomalies even in the absence of prior knowledge about specific anomaly types.

In this work we develop a probabilistic approach for modelling dynamic networks with anomalous edges, termed \dmacd, which stands for Dynamic model for Anomaly, and Community Detection, where nodes remain fixed, and edges appear and disappear. Nodes have community memberships that determine the regular behavior of edge formation over time. The model estimates both community structure and the probability that an edge is anomalous.

We evaluate the effectiveness of our approach on real-world datasets and demonstrate its ability to accurately detect anomalous events in dynamic networks. By studying a dataset of players transfers between professional men football clubs, we find a variety of anomalous behaviors, ranging from  anomalies appearing between clubs with compatible community structure but unusual frequency of transfers, or between clubs that have incompatible communities. Finally, we illustrate a potential application relevant for data collection procedures that require an initial preprocessing steps, showing how our model can be used to flag potential errors caused by mislabeling of nodes.

\section{Model: \dmacd }
\label{sec:model}
We observe pairwise interactions between individual nodes over time, which we represent using a time-dependent adjacency matrix $A_{ij}(t)$. In this work, we focus on binary directed temporal networks, meaning that $A_{ij}(t) =1,0$ indicates the existence of an edge from node $i$ to $j$ at time step $t$, or not, respectively.

To model the structure of temporal networks, we adopt a continuous-time Markov process approach similar to \cite{Safdaridyn_2022, zhang2017random}. We assume that the appearance and disappearance of edges are continuous events that occur on real-valued timescales. However, since our observations are discrete, we track the network at discrete time steps, $t=0,1,...,T$.

We define a binary variable (or label) $Z_{ij} \in \ccup{0,1}$ to classify anomalous edges: if $Z_{ij}=1$, then the pair of edges $(i,j),(j,i)$ is anomalous, it is regular otherwise. This is a static quantity, as in this work, we focus on the scenarios where the status (or label) of a pair of nodes is either anomalous or not. We assume this not to change in time, while edges can appear and disappear based on their label. We also consider it to be symmetric, i.e., $Z_{ij}=Z_{ji}$, as we assume that an anomaly involves a given pair of nodes, regardless the direction of an edge (but $A_{ij}(t)$ is in general asymmetric, to model more generally directed networks). This label $Z_{ij}$ is not known a priori, it is a latent variable that we need to learn from data, together with the other parameters of the model.

With this in mind, we  express the likelihood of the existence of the edge $(i,j)$ at time step $t=0$, given a set of latent variables $\Theta$ as:
\be
P (A_{ij}(0)|\Theta ) =  \pois(A_{ij}(0); \lambda_{ij}(0))^{1-Z_{ij}} \,\pois(A_{ij}(0); \pi)^{Z_{ij}} \;,\label{eqn:A0} 
\ee
where $\pi$ is a constant and  the hyper-parameter $\mu \in [0,1]$ regulates the prior distribution of $Z$, 
\be
Z_{ij}  \sim \bern(\mu)\;.\label{eqn:Zprior}
\ee 

In other words, at an initial time step $t=0$, a Poisson distribution with mean $\pi$ governs  the probability of an anomalous edge at the initial time step. Instead, the existence of a regular edge between nodes $i,j$ is dictated by a Poisson distribution with mean $ \lambda_{ij}(0)$. This parameter depends only on the communities to which the nodes belong, according to a mixed-membership model \cite{de2017community}:
\be\label{eq:rappear}
\lambda_{ij}(0)  = \sum_{k,q}u_{ik}v_{jq}\,w_{kq}(0)  \quad.
\ee   
The $K$-dimensional vectors $u_i$ and $v_i$ represent the out-going and in-coming communities of nodes $i$, respectively. The $K \times K$ affinity matrix, with entries $w_{kq}(0)$ plays a crucial role in regulating the community structure of the network. When the diagonal entries exceed the off-diagonal entries, it promotes assortativity and increases the likelihood of edges connecting nodes within the same community. The number of communities $K$ is a hyper-parameter that needs to be selected with model selection criteria. In our experiments here we use 5-fold cross-validation.\\
Given the probability of the network at the initial time step, and the labels of edges $Z_{ij}$, we now proceed with the evolution in time of $A_{ij}(t)$, $\forall t=1,\dots,T$. We consider a Markov process to approximate the trajectory in time as:
\begin{small} 
\be
P(\{A(t)\}| \Theta )  =\prod_{i,j} \Bigg\{  P \left(A_{ij}(0)|\Theta \right)    \times \prod _{t=1}^T \{P(A_{ij}(t)|A_{ij}(t-1),\Theta )\} \Bigg\} \quad .    \\  \nonumber
\label{eq:joint_dist}\ee
\end{small}\\
The dynamic is determined by the transition probability, that we assume factorized as
\be
P(A_{ij}(t)|A_{ij}(t-1),\Theta)=\rup{P_{a}(A_{ij}(t)|A_{ij}(t-1),\Theta)}^{Z_{ij}} \times \rup{P_{r}(A_{ij}(t)|A_{ij}(t-1),\Theta)}^{1-Z_{ij}} \quad,
\ee
where the subscripts $a,r$ denote the evolution of an anomalous or regular edge, respectively. In this work, we assume that
at each time-step, regular  edges appear with rate $\lambda_{ij}(t)= \sum_{k,q}u_{ik}v_{jq}\,w_{kq} (t)$, and disappear with rate $\gamma$, while anomalous edges appear and disappear with the rates $\ell$ and $\zeta$, respectively.  Hence, to fully build the time evolving probability of the node pairs, it becomes essential to estimate the transition rates that govern the appearance  and disappearance of both regular and anomalous edges over time.\\
In line with the methodology employed by Zhang et al. \cite{zhang2017random, SafdariJBD2022}, we adopt a similar approach to compute the probability of edge existence. This involves solving a master equation that governs the evolution of the presence of edges, including both regular and anomalous ones (refer to \Cref{appendix:model_development,appendix:master_eq}  for more details).\\
By solving the master equation for the edge probability outlined in Eq.(\ref{eqSI:mst1}), the transition rates specified in Eq.~(\ref{eqSI:TrasitionRates}) can be derived based on the model parameters denoted by $\Theta \equiv \{ u, v, w,\beta, \mu, \pi,\phi,\ell \}$, see Eq.~(\ref{eqSI:TransProb}). Hence, we are equipped to finalize the posterior computation for the edges, as articulated in Eq.~(\ref{eqSI:Lconv0}).\\
The goal of our study is to learn the hidden parameters of the model, $\Theta$, based on the adjacency matrices observed at each time step $\ccup{A(t)}_{t=0}^{T}$. To accomplish this objective, we perform an inference task by maximizing the log-posterior of the parameters, $\mathcal{L}(T, \Theta)$,  given the data (refer to \Cref{appendix:sec:inference} for additional details). \\
An important quantity that we are able to estimate in closed-form is the the posterior distribution of the labels $P(Z_{ij}|\ccup{A_{ij}(t)}_{t=0}^{T},\Theta)$, and in particular the expected value $\mathbb{E}[Z_{ij}]=Q_{ij}\in [0,1]$. Other methods, e.g., approaches based on deep learning and embeddings, do not have a straightforward way to estimate this; hence, they need to make use of ad-hoc choices to classify anomalies a posteriori using proxies for this quantities (see \Cref{appendix:sec:embeddings} for a detailed discussion). In our case, the $Z_{ij}$ are encoded explicitly in our model.\\
We made some main assumptions about the temporal dynamics of the model parameters. We consider the affinity matrix as a variable dependent on time, while keeping the community membership vectors $u_i$ and $v_i$ static over time.  It is worth noting that an alternative perspective can be achieved by keeping $w$ fixed and permitting changes in the community membership vectors over time \cite{matias2015statistical}. Either of these approaches allow the community structure to evolve dynamically impacting edge formation in time via the parameter $\lambda_{ij}(t)$.
Our model assumes a fixed number of communities $K$, yet employs a mixed-membership model to assign nodes to multiple communities with varying intensities. This allows to effectively capture the likelihood of the data by adjusting how an entry $u_{ik}$ or $v_{ik}$ impacts the magnitude of $\lambda_{ij}(t)$ via $w(t)$, while maintaining $K$ constant.
An overview of the algorithm to estimate the hidden parameters of the model is described in \Cref{alg:EM}.

\input{algo_dyn_acd} 

\subsection{Interpretatibilty} 
The probabilistic framework and the explicit incorporation of latent variables in  \dmacd\ facilitate a clearer understanding of the factors driving community detection and anomaly detection, as we discuss in more detail here.\\
First, unlike black-box methods, \dmacd\ is based on a generative model where the mechanism through which community structure and anomalies influence edge formation over time is specified explicitly. By understanding the underlying generative process, practitioners can gain insights into why certain edges are flagged as anomalous. We illustrate examples of different interpretations in the analysis of football players transfer below, where we see how anomalies arise between nodes that have compatible or incompatible communities structure, and how the time dynamics of edge formation can also play a role.\\
Second, the model estimates latent variables, including community memberships ($u_i$ and $v_i$) and anomaly indicators ($Z_{ij}$). This explicit representation provides a clear rationale for edge formation patterns. For instance, in our study of football players transfers, we observe that top clubs have mixed memberships, indicating their broad trading patterns across different markets. This contrasts with smaller clubs, which typically have more concentrated community memberships. This nuanced understanding is possible due to the explicit estimation of these latent variables.\\
Third, \dmacd\ allows for closed-form estimation of the posterior distribution of anomaly labels, $P(Z_{ij}|\{A_{ij}(t)\}_{t=0}^{T},\Theta)$, and their expected values, $\mathbb{E}[Z_{ij}] = Q_{ij}$. This is a significant advantage over other methods that often rely on ad-hoc measures to classify anomalies. The ability to compute $Q_{ij}$ provides direct interpretability by quantifying the likelihood of each edge being anomalous. Deep learning methods that are based on node embeddings, require making arbitrary decisions on how the inferred embedding determine the probability of an edge being anomalous (e.g. using the Hadamard product of two node embeddings).

\subsection{Computational Complexity and Limitations} 
The computational complexity of the EM algorithm in \dmacd\ is dominated by a term $ O(N^2)$, necessary to compute the matrix $\boldsymbol Q$ in Eq. (\ref{appendix:eqn:Qij}), and this is in general a dense matrix. While the matrix $\boldsymbol Q$ is essential for the model's ability to detect anomalies, its presence could make the algorithm less feasible for large-scale networks, particularly as the number of nodes $N$ grows. The $O(N^2)$ complexity represents the worse-case scenario. One could in principle reduce this by using efficient ways to compute quantities that require estimates of $Q$ without having to store and compute all the $N^2$ entries at any given time, but only a subset of them. This could be done for instance by smart ways of caching values from one algorithmic iteration to the next. In this work we have not explored these possibilities.

A factor that may limit the inference performance is data sparsity. Estimating parameters is usually more difficult in sparse datasets, where the number of observed edges is small relative to the system size. This could be a problem in particular in dynamic networks, where there could be various ways to aggregate edges into time steps, thus impacting the sparsity of the dataset given in input to our algorithm. Depending on the reference time window considered to denote a particular time step, one could observe too little information, when the window is too short, or too much to lose the dynamic character, when the window is too long. For instance, in the Transfermarkt data below we find that a four-year interval provides a more stable and temporally meaningful representation of the dataset, allowing us to capture significant trends and patterns over time without being hindered by the variability and sparsity that a yearly division would introduce. An interesting direction for future work would be to device principled ways to select an appropriate time window that is most informative in terms of detecting anomalies. We do not explore this here.\\

\section{\label{sec:res} Results} 
\subsection{\label{sec:res_syn} Synthetic Networks} 
\begin{figure}[t] 
  \includegraphics[width=0.5\linewidth]{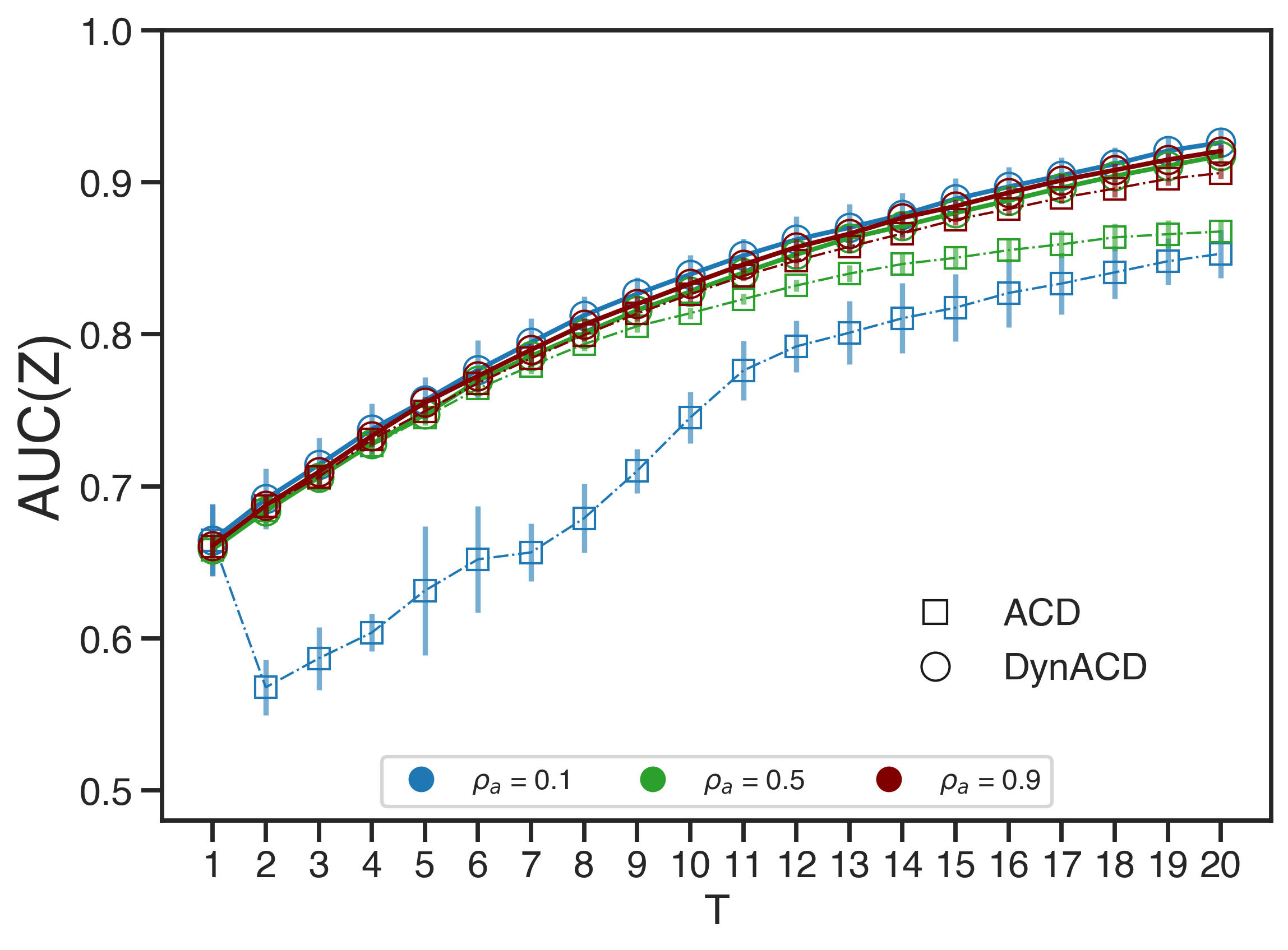}  
  \caption{\textbf{Anomaly detection in synthetic networks.} The AUC(Z) metric quantifies the model's ability to distinguish between regular and anomalous edges.  The synthetic network has $N = 300$ nodes, average degree $\langle k \rangle=8$, and $K = 8$ communities of equal-size unmixed group membership generated with our generative model.  Here,  $\beta=0.2$, $\ell=0.2$, $\phi=0.2$.  Lines are averages and standard deviations over $10$ sampled networks.  \dmacd\  ($\circ$), and  \acd \text{}  on aggregated dataset (\protect\scalebox{0.75}{$\square$}). }
  \label{fig:AUCZall}
\end{figure}

As the primary objective of this study is to develop an algorithm for anomaly detection, it is essential to evaluate the predictive capability of the \dmacd\  algorithm in identifying anomalous edges over time.  
To this end, we generate synthetic networks, using our model, and evaluate its performance in detecting anomalies in a situation where we know the ground truth anomalies. These are non-zero entries of the matrix $Z$ used to generate the data.
We utilize the matrix $Q$ with entries $Q_{ij}$ (refer to \cref{appendix:eqn:qQij}), representing the model's estimations of anomaly score for each edge.   We consider  the area under the receiver operating curve (AUC) as prediction performance evaluation metric. A higher AUC score signifies enhanced performance in accurately identifying true anomalies.  We are interested in particular in assessing how performance is enhanced by exploiting the dynamic character of the data. To this end, we compare the performance of \dmacd\ against an algorithm developed for anomaly detection in static networks, \acd\ \citep{SafdariJBD2022}.  \acd \ does not accept dynamical data as input. Instead, at each time step, we apply \acd\ to the aggregated dataset, i.e., the sum over the time dimension. As a consequence, \acd\ predicts anomalies in a dataset where the temporal information is lost. \\
In both cases, we observe a discernible increase in the model's anomaly detection capabilities as we increase the number of time steps, $T$,  which is given as input to the algorithm, as shown in \Cref{fig:AUCZall}. However, \dmacd\ achieves significantly higher AUC values when the density of anomalies is low ($\rho_{a}=0.1$); hence there are few positive examples of anomalies for a model to detect. For a balanced density ($\rho_{a}=0.5$), performance is similar for a low number of time steps, however, subsequently the static model tends to plateau, while \dmacd\ keeps increasing AUC,  as more time steps are given in the  input. This behavior is further observed, although to a lesser extent, for more extreme density of anomalies ($\rho_{a}=0.9$). These results highlight the advantage of using the temporal information. When density of anomalies is low, the dynamical model can better recover them. When density is higher, the dynamical model better benefits of an increasing number of time steps given in input. \\ 
 Further quantitative examination of \dmacd\ and comparison with \dmcd \ for synthetic networks in additional settings can be found in \Cref{appendix:sec:res_syn}. 

%----------------------------------------------------------------------------

%____________________________________________________________________________________

\subsection{Detecting anomalies in real world dynamical networks}\label{sec:res_rd} 
We study temporal real-world networked datasets characterized by a wide spectrum of nodes and edges, originating from diverse contexts such as trading (Transfermarkt), transportation (US-Air), citation (SCC2016), communication (Eu-monthly), and social networks (UCI Messages), see \Cref{appendix:data_desc} for a detailed overview.    \\
Our goal is to detect anomalous edges in these different datasets. As in these cases, the true labels for anomalous or regular edges are unavailable, we  manually inject  random edges into the dataset \citep{Akoglu2015,YuNetWalk2018,Safdari2022ACD} and label them as anomalous.  Then, we run our anomaly detection algorithm and evaluate the model's ability to identify anomalies. 
Specifically, we inject $n$ random edges between nodes within a given anomaly density range, $\rho_a$. Afterwards, we apply our algorithm to the modified dataset, calculating the expected value $\mathbb{E}[Z_{ij}]=Q_{ij}$ for the edge labels, i.e., the probability that the edges between two nodes $i,j$ are anomalous. Following this step, we assign labels to the edges. In this particular experiment, we identify the initial $n$ pairs ($i,j$) with the highest $Q_{ij}$ values as anomalous edges. As a performance metric, we calculate the recall, which measures the accuracy of the algorithm in correctly identifying True Positives; and the AUC, the ability to rank True anomalous edges higher than False anomalous ones. Higher values mean better classification performance in both cases. Notice that with this procedure, as we fix the number of detected anomalous edges equal to the number $n$ of injected, we have that recall is equal to precision.\\

\begin{figure}[t] 
\includegraphics[width=0.95\linewidth]{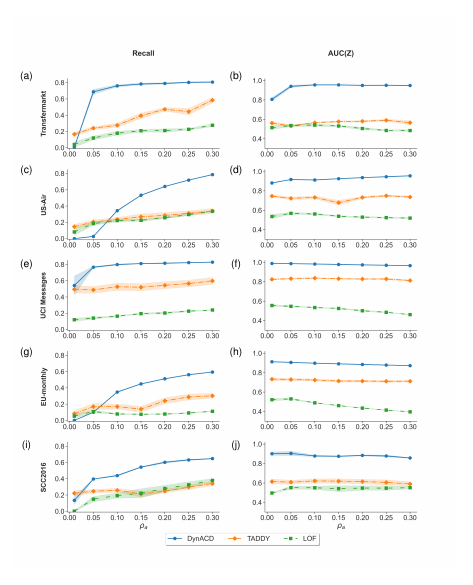}   
\caption{\textbf{Anomaly Detection in Real-World Datasets.} Comparison of the performance of \dmacd, TADDY, and LOF in detecting anomalies across five real-world datasets. Each row corresponds to a dataset, with the left column showing the recall and the right column showing the AUC(Z) scores as the fraction of injected anomalies $\rho_{a}$ increases. \dmacd\ (solid lines) demonstrates a significant improvement in both recall and AUC(Z), outperforming TADDY (dash-dotted lines) and LOF (dashed lines). LOF shows relatively poor performance, with AUC(Z) values close to random guessing and low recall scores, particularly at higher $\rho_{a}$ levels. Lines and shaded areas represent the mean and standard deviation over 10 sampled networks, respectively.}
\label{fig:Ensemble_injection} 
\end{figure}

We find a robust performance of \dmacd\ in detecting anomalies across the five datasets, as shown in \Cref{fig:Ensemble_injection}. We observe that recall increases with the fraction of injected anomalies $\rho_{a}$, reaching a plateau at around $0.8$, depending on the dataset, \Cref{fig:Ensemble_injection}(a,c,e). Similarly, we find high values of AUC(Z) in all datasets, with values consistently above $0.85$, \Cref{fig:Ensemble_injection}(b,d), even in cases where the AUC(Z) shows a decreasing trend with increasing $\rho_{a}$, \Cref{fig:Ensemble_injection}(f,h,j). These results confirm the robustness of \dmacd\ in detecting anomalous edges in different scenarios.

We compare the performance of \dmacd\ against two baseline methods: TADDY \citep{liu2021anomaly} which uses  transformers for anomaly detection in dynamic graphs, and the Local Outlier Factor (LOF) algorithm \cite{breunig2000lof}, with the implementation available as part of the python package \texttt{scikit-learn} \cite{scikit-learn}. TADDY leverages the attention mechanism in transformers to identify anomalies by capturing temporal dependencies and node interactions. LOF is a density-based method that detects anomalies by evaluating the local deviation of an edge's density compared to its neighbors, using the structural and temporal features described above.

As illustrated in \Cref{fig:Ensemble_injection}, \dmacd\ consistently outperforms both TADDY and LOF across all datasets. In the recall plots (left column), for very small anomaly densities ($\rho_{a} < 0.05$), \dmacd\ has comparable results to TADDY and LOF. However, for $\rho_{a}$ values greater than $0.05$, \dmacd\ significantly outperforms both baselines, achieving higher recall scores. TADDY's recall is generally higher than that of LOF, but both methods perform notably worse than \dmacd, particularly at higher $\rho_{a}$ levels.

In the AUC(Z) plots (right column), \dmacd\ consistently achieves higher values, exceeding $0.85$ across all datasets and anomaly ratios. TADDY shows lower and more variable AUC(Z) values compared to \dmacd, while LOF's AUC(Z) values are close to those based on random guessing (AUC of $0.5$), indicating poor performance. This suggests that LOF does not perform significantly better than a random classifier in this context.

The superior performance of \dmacd\ can be attributed to its ability to effectively capture both the dynamic and topological features of the graphs, which are essential for identifying anomalous edges. In contrast, LOF, which relies on static features and local density estimation, may not fully capture the complex temporal dynamics of the graphs, leading to suboptimal performance. TADDY, while designed for dynamic graphs, may not generalize well across different datasets or may be limited by its model capacity.
Lines and shaded areas in \Cref{fig:Ensemble_injection} represent the mean and standard deviation over 10 sampled networks, respectively, demonstrating the consistency of our method's performance.

%%%
%%%

\subsection{Analysis of football player transfers from Transfermarkt}
\label{sec:res:Transfermarkt}  

To illustrate the applicability of our model, we run the algorithm on the Transfermarkt dataset, which contains information on player transfers among European men's football clubs. The dataset contains transfers involving eight different professional men's football leagues in Europe, see \Cref{appendix:data_desc}. Here nodes are clubs and directed edges $i\rightarrow j$ represent transfers from club $i$ to club $j$. We focus on transfers made during the period from season 2008/2009 to 2022/2023. We have divided the dataset into five distinct time snapshots, each encompassing a four-year duration. The choice of aggregating the dataset into four-year snapshots was made to preserve sufficient data density and incorporate meaningful temporal pattern.  We consider a network where $A_{ij}^{t}=1,0$ if there is at least a player transferred from $i$ to $j$ in period $t$ or not. 

To enrich the qualitative interpretation  of our results, we utilize metadata for club categorization. This includes the transaction volume associated with player transfers into and out of clubs, representing a proxy for their financial wealth. Overall, we have a total of $N=261$ clubs across eight leagues and $E=\sum_{i,j,t}A_{ij}^{t} = 7613$  transactions.   The number of communities in the network is set to $8$, which is inferred from  5-fold cross-validation.\\

\subsubsection{Community analysis}
\begin{figure}[h] 
\includegraphics[width=1. \linewidth]{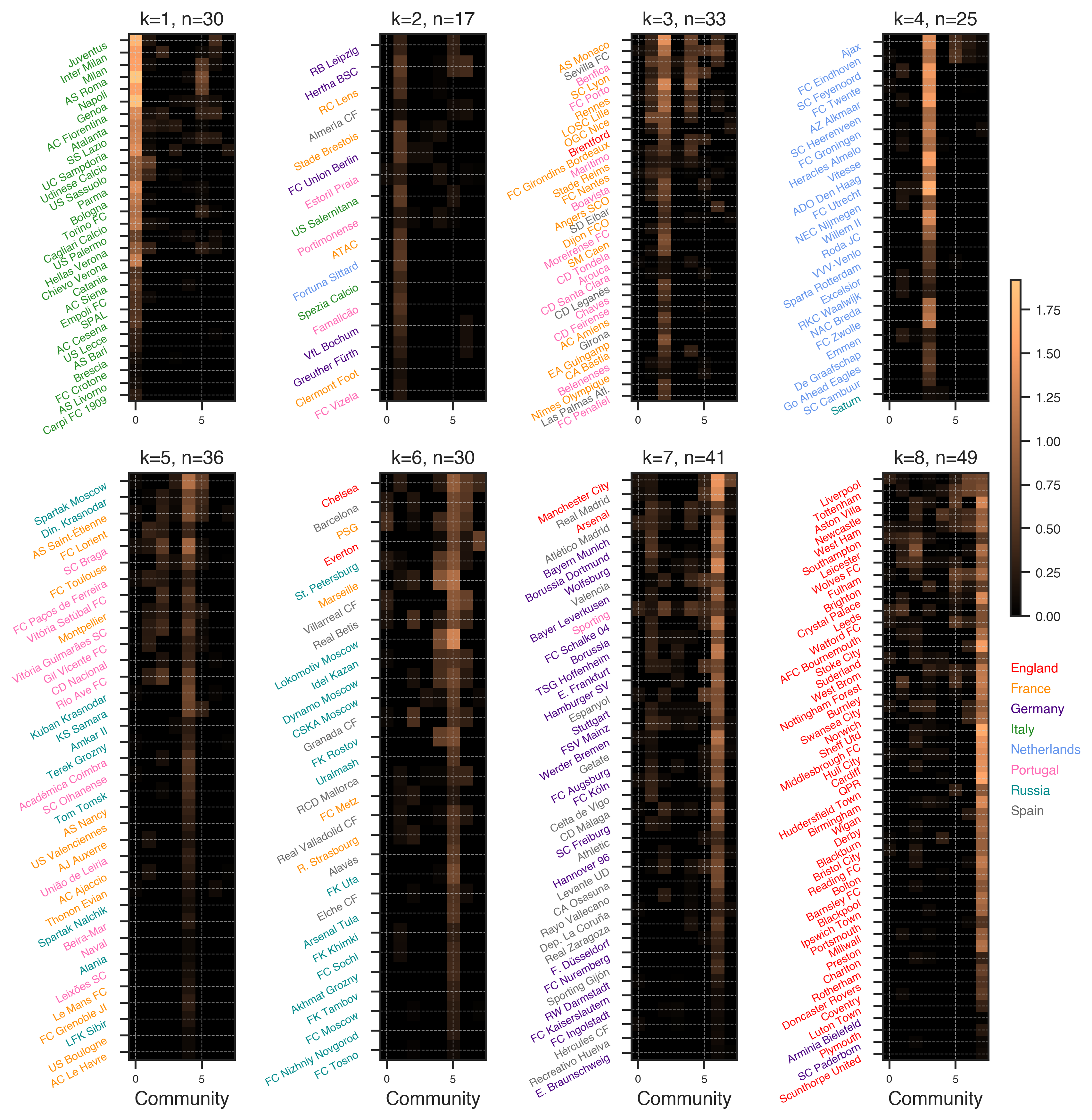}   
\caption{\textbf{Communities in the Transfermarkt dataset:} In-coming (soft) community membership $v$ of clubs. The colors of the y-labels indicate the country to which the receiving clubs belong. This plot reveals an alignment between the community membership of clubs and their respective nationalities. The corresponding country of each league is shown in the legend on the right with the color assigned to that league.}
\label{fig:Transfermarket_community_v} 
\end{figure}

Our algorithm assumes that community membership is a primary driver of player transfers between football clubs,   with the idea that one can identify main patterns of behaviors and thus cluster clubs in groups. These patterns can be influenced by different factors. For instance, we expect that the country of a club plays an important role, as this shapes the main market in which the club is embedded in.  
By analyzing the distribution of community membership we see that indeed country correlates well with the inferred partitions, see \cref{fig:Transfermarket_community_v} for an example of the in-coming groups based on $v$. Indeed 4 out of 8 inferred groups are made mainly by clubs in the same countries (Italy, Netherlands, England, and Russia). Other groups are more mixed, with a group made mainly of German and Spanish clubs. French and Portuguese clubs are split into two main communities and, to a lesser extent, into other groups. In one community, the country does not seem to play a role. It is a small community of 17 clubs that have a low magnitude in all the entries of $v_{i}$. These are small clubs that typically classify in the bottom part of the league table, have a low wealth category, and do not trade much in general. The main exception is RB Leipzig, which is a top European club. We argue that it is classified in this group because this dataset does not fully capture the pattern of player transfers that concerns this particular club. The Transfermarkt dataset contains only deals that involve clubs in either of the eight considered leagues. However, RB Leipzig is very active in markets outside these  eight countries, hence our data may be limited in capturing the behavior of this club.  Similar patterns are observed for the out-going communities based on $u$, see \cref{figSI:Transfermarket_community_u}. \\Finally, one can notice how top clubs tend to have a more mixed-membership as they trade more broadly across different markets, and their in-coming and out-going membership differs. For instance, Manchester City and Chelsea have in-coming communities that differ from those of the majority of English clubs, while their out-going ones are more aligned with their country. This signals that, while they tend to sell players more often to other English clubs, they tend to instead acquire players from other markets.

\subsubsection{Club wealth and anomalous edges}
\begin{figure}[p] 
\includegraphics[width=1 \linewidth]{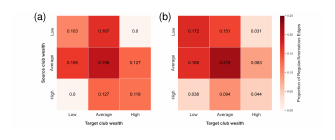}     
\caption{\textbf{Transfermarkt datasets: Edges Distribution by wealth category of clubs.}  
Each cell in the heatmap represents the proportion of edges between clubs, distinguished by wealth categories (High, Average, and Low). We separate the set of edges considered in each heatmap between a) anomalous and b) regular, as estimated by \dmacd. We normalize the heatmaps by the total number of edges considered in each plot, so that the sum over all entries in each heatmap equals 1. Darker color means higher proportion of edges exchanged between clubs in the given wealth categories.} 
\label{fig:Transfermarket_heatmap}  
\end{figure}

In addition to communities, our algorithms naturally output estimates of anomalous edges. We analyze how these edges are distributed between clubs based on their wealth. We extract three tiers based on the 
total transfer fees spent in the years covered by the dataset by each club. Specifically, we label a club's wealth as ``high'' when this number is higher than 800 M\EURtm; ``low'' if is lower than 150 M\EURtm; and ``average'' if it is in between these two extremes. It should be noted that the thresholds for each category were determined based on the observed distribution of the clubs' wealth. Then, we measure how regular and anomalous edges are distributed based on the wealth of the in-coming and out-going club. In \cref{fig:Transfermarket_heatmap},  we notice a significantly higher fraction of anomalous edges involving wealthier clubs (``high'') at both ends. Specifically, compared to the clubs engaged in regular edges, there is a notable increase in the number of transfers from wealthier clubs to both wealthier  ($0.119$ for anomalous edges vs. $0.044$ for regular edges),  and average wealth ($0.127$ for anomalous edges vs. $0.094$ for regular edges) clubs.  Conversely, we notice a higher fraction of transfers between low wealth clubs in regular edges, suggesting that is less common to observe an anomalous edge between clubs of lower wealth.

\subsubsection{Interplay between anomalous edges and communities} 
In the previous paragraphs, we focused on describing communities and anomalies separately. However, our model combines them to explain the observed transactions in time between clubs. Hence, here we investigate how these two factors together contribute to explaining edge formation, and illustrate various concrete scenarios where their interplay can guide practitioners in their analysis.

\subsubsection*{Anomalies between clubs with compatible communities}
\begin{figure}[h] 
\includegraphics[width=1.05 \linewidth]{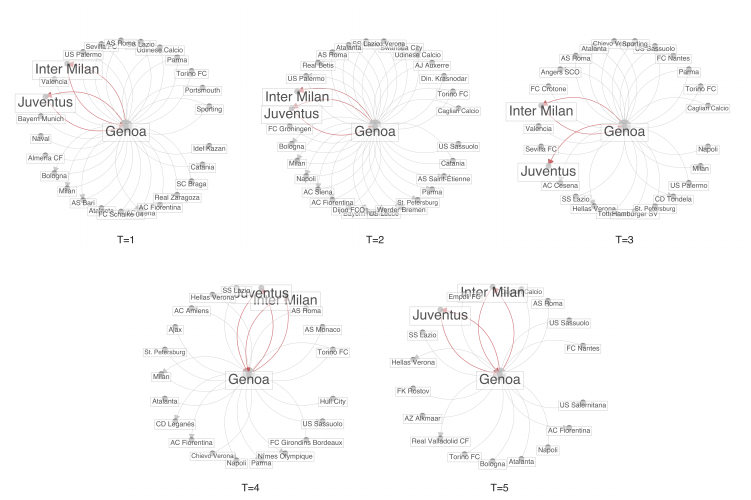}    
\caption{\textbf{Transfermarkt datasets:  Genoa transfer network}:  Visualization of player transfers to and from Genoa involving  various clubs at different time steps.  Notably, there is a consistent presence of transfers with  Juventus and Inter Milan at most time steps.}
\label{fig:Transfermarket_genoa} 
\end{figure}

The previous analysis studied coarse-grained patterns of anomalies based on club wealth. Having access to individual estimates of $Q_{ij}$, the expected value of an edge to be anomalous, we can also isolate fine-grain patterns described by the behavior of individual clubs. One such pattern is that involving clubs in the same community. While we would usually expect more edges to exist between nodes with higher $\lambda_{ij}^{t}$, as dictated by their community memberships, we can still observe high $Q_{ij}$. This is because anomalies may also stem from unusual dynamical patterns, i.e., unusual distribution in time of transfers than we would otherwise expect between clubs with high  $\lambda_{ij}^{t}$. One example of such a case is that of transfers involving pairs of Italian clubs, Genoa and Juventus and Genoa and Inter Milan, as shown in \Cref{tab:Genoa_lambdaQ}. Comparing the distribution of transfers in time between these clubs and those involving other pairs with similar community memberships and $\lambda_{ij}^{t}$, we can attribute anomalies to the frequent appearance and disappearance of transfers between these specific clubs, impacting the expected value of the anomalous parameter, as detailed in Eqs.~(\ref{appendix:eqn:Qij}-\ref{appendix:eqn:F-funcs}). We notice in fact that, while we expect transfers between Genoa and Juventus or Inter Milan to happen, we observe a higher number than expected, in both directions over time. In other words, it is not the presence of transfers between these clubs that is anomalous, but their frequency.  This can also be seen in \Cref{fig:Transfermarket_genoa}, where we isolate all the transfers involving Genoa and notice how Juventus and  Inter Milan are the only Italian clubs that exchange players with them consistently across time.
This pattern is inherent to the dynamic nature of the data and could not be naturally captured by static models for anomaly detection. 
\\
\subsubsection*{Anomalies between clubs with incompatible communities} A different pattern of anomaly is that involving clubs with low $\lambda_{ij}^{t}$. When several edges exist between pairs of clubs in a way that is not expected by their community memberships (as measured by a low $\lambda_{ij}^{t}$), our model flags also these types of transfers as anomalous. A notable example is a set of edges involving Udinese and Watford, an Italian and an English club, respectively. Both clubs have memberships $u_{i}$ and $v_{i}$ peaked in one community (the one of mainly Italian and mainly English clubs, respectively). From this estimate, we expect the two clubs to transfer players mainly with other clubs in the same country. However, we observe an unusual high volume of transfers between these two clubs, something that is flagged as anomalous by our model, see \Cref{tab:UdineseWatford_lambdaQ}. As the total number of transfers ($A_{ij}(t)=7$) is not an anomaly in general (it is observed in several other pairs of clubs that are estimated as regular), we can attribute the high estimated $Q_{ij}$ to the fact that it is a high number for clubs in these two communities specifically. We should remark that Udinese and Watford shared ownership in the years tracked by the dataset, which could explain the unusually high number of transactions. Note that this information was not given in input to the algorithm. Instead, observing a high value of $Q_{ij}$ prompted further investigation into the relationship between these two clubs, which illustrates a practical usage of this model. \\
A more extreme case under this scenario is when some edges are observed between clubs with $\lambda_{ij}^{t}=0$. Depending on what time step the edges are observed, this could imply the extreme value $Q_{ij}=1$, see Eq.~\ref{appendix:eqn:Qij}. In this case, it may be enough to observe one player transferred between two clubs, which have strongly mismatched community memberships, to flag the edge as anomalous, for instance transfers between NEC Nijmegen and CA Osasuna in \Cref{tab:UdineseWatford_lambdaQ}.\\ 
\subsubsection*{Case study: mislabeling in data collection.} Having discussed different types of anomalies captured by our model, we now illustrate relevant case scenarios that apply more broadly when collecting a dataset. For this, we manipulate the dataset by merging the clubs AC Milan and Inter Milan, two distinct Italian clubs with similar names. The high similarity in their name can cause confusion in data collection processes that start with a preprocessing step to clean node names to identify them uniquely. For instance, it is common to have several repetitions of the same node with slightly different names, due to errors or imprecisions in the data collection, particularly when node names come as string types (as opposed to integer numbers). This was a necessary step also in the Transfermarkt data, where many clubs appeared multiple times with slightly different string names. In our example, ``Milan'', ``A.C. Milan'', ``AC Milan'', and ``Milan AC'' refer all to the same club, whereas ``Inter Milan'', ``FC Internazionale'', ``Inter'' refer to another club; but ``Inter Milan'' can cause confusion. In our experiment, we merge the two clubs simulating a possible error caused by mislabeling clubs. As a result, we obtain one node to represent two clubs, and players transferred involving either of these two clubs are all assigned to that unique node ``Milan''. All the rest of the dataset is kept the same as before. We run the model on this manipulated data and obtain several anomalous edges as in the first case described above, i.e.,  high $\lambda_{ij}^{t}$ and high $Q_{ij}$, as shown in \Cref{tabSI:Milan_InterMilan}. All these edges involve the ``Milan'' node and several other clubs (all Italian but Manchester City). The anomalies stem from the high frequency of transfers (in several cases $A_{ij}^{t}=T$, $\forall t=1,\dots,T$), which is higher than what is expected by communities alone. This high frequency of anomalies involving one particular club prompts practitioners to investigate further, and likely to identify the data collection problem.

% --------------------------------------- Table Genoa  --------------------------------------------------------------------------
\begin{table}[H]
\caption{\bf {Transfermarkt dataset:  frequency of transfers between Genoa and Juventus or Inter Milan}. Illustrative instances of player transfers between the clubs, characterized by a high expected value for tie formation based on community membership, here measured by a high average $\langle \lambda_{ij}^{t}\rangle=\sum_{t}\lambda_{ij}^{t} / T$. However, these interactions were flagged as anomalous (high $Q_{ij}$) due to higher rates of appearance.}
\begin{center}
\begin{ruledtabular}  
\begin{tabular}{llcccccccc}
Source Club ($i$)&      Target  Club  ($j$)&  $A_{ij}(0)$ &  $A_{ij}(1)$ &  $A_{ij}(2)$ &  $A_{ij}(3)$ &  $A_{ij}(4)$ &  $\sum_{t}A_{ij}(t)$ &  Q$_{ij}$ &  $\langle \lambda_{ij}^{t}\rangle$\\ \hline \\
    Genoa &    Juventus &          1 &          1 &          1 &          1 &          1 &      5 & 0.425 &      0.376 \\
   Juventus &       Genoa &          1 &          1 &          0 &          1 &          1 &      4 & 0.425 &      0.483 \\
   \hline \\
      Genoa & Inter Milan &          1 &          1 &          1 &          1 &          1 &      5 & 0.357 &      0.340 \\
Inter Milan &       Genoa &          1 &          0 &          1 &          1 &          1 &      4 & 0.357 &      0.427 \\
\end{tabular} 
\end{ruledtabular} 
\end{center}
\label{tab:Genoa_lambdaQ}
\end{table}
% ---------------------------------------

% ---------------------------------------Figure RD ---------------------------------------
% ---------------------------------------

% ---------------------------------------Figure RD ---------------------------------------  
% ---------------------------------------

% ---------------------------------------Figure RD ---------------------------------------  
% ---------------------------------------  

\clearpage
\section{Discussion}\label{sec:discussion}
 
We have developed a principled  probabilistic generative model that  integrates mechanisms for the emergence of anomalous edges within a dynamic network. The model's core is rooted in the community memberships of nodes as the structural parameters that determine regular patterns. The network interactions governed by this mechanism are identified as the regular edges in the network. Instead, deviations for regularity are detected as anomalous.\\
One of the key advantages of our approach lies in its ability to provide targeted anomaly detection within specific communities of the network. Traditional anomaly detection methods often struggle to localize anomalies, treating the entire network homogeneously. In contrast, \dmacd\  exploits the localized nature of anomalies, recognizing that deviations from regular behavior often occur within specific subsets of the network. This targeted approach enhances the accuracy and efficiency of anomaly detection, particularly in scenarios where anomalies manifest in localized regions rather than affecting the entire network.\\
Our approach models evolving network structures, allowing us to detect anomalies without prior knowledge. Empirical tests show the effectiveness of \dmacd\ in spotting anomalous events in dynamic networks, particularly when the temporal aspect of the data is essential for understanding network structure.\\
Furthermore, the application of \dmacd\   to the analysis of football player transfers illustrates its practical utility in real-world scenarios. By uncovering anomalous transfer patterns between football clubs, \dmacd\   highlights potential irregularities that may signify mislabeling in data collection processes or unusual dynamics in player interactions. This case study highlights the importance of integrating domain knowledge with data-driven anomaly detection methods, which allows identifying and interpreting anomalies in complex networks. \\ 
Notably,  our model is formulated on multiple assumptions. First, we assume community structure to capture regular patterns. Hence, our model may not be appropriate to model datasets where a community structure is not expected. Second, we assume that anomalies are static: although edges can appear and disappear in time their regular or anomalous nature does not change. Also in this case, if we expect an edge to change in nature from regular to anomalous, or vice versa, our model may not be appropriate. One can relax this assumption, for instance by making the variables $Z_{ij}$ time-dependent. This would require further specifying a mechanism for their transition in time and an increased number of parameters, potentially leading to over-fitting and higher algorithmic complexity. Third, we assume a Markovian transition probability for edges to appear and disappear. If we expect significant time-correlations between past history beyond that one-step away, then we would need a model that is able to incorporate memory into the system.\\ 
While we have promising results in confirming the reliability and applicability of  the proposed algorithm in real world scenarios,  there are still several directions to extend the model.  Integrating topological properties as reciprocity \citep{safdari2020generative,contisciani2022community} or triadic closure \cite{peixoto2022disentangling} could enhance its expressiveness in representation of real-world interactions. 
Exploring how to integrating both node and edge anomalies is a natural future direction, considering their coexistence in datasets. This could be potentially addressed by also incorporating extra information as node attributes \cite{contisciani2020community,newman2016structure,fajardo2022node}. In general, main challenges remain addressing the scalability of these method to handle large datasets and the ability to incorporate the intuition of experts and provide them interpretable results.
\newpage

\section*{Data Availability}
The synthetic datasets generated in this study and the corresponding codes can be found in  \url{https://github.com/hds-safdari/DynACD}.

\section*{Code Availability}
We provide an open source implementation of the code online at \url{https://github.com/hds-safdari/DynACD}.

\section*{Acknowledgements}
\vspace{-0.1in}
{All the authors were supported by the Cyber Valley Research Fund.

\newcommand{\beginsupplement}{%
        \setcounter{table}{0}
        \renewcommand{\thetable}{S\arabic{table}}%
        \setcounter{figure}{0}
        \renewcommand{\thefigure}{S\arabic{figure}}%
        \setcounter{equation}{0}
        \renewcommand{\theequation}{S\arabic{equation}}
         \setcounter{section}{0}
        \renewcommand{\thesection}{S\arabic{section}}
 }

%\beginsupplement 
%\appendix
%\section*{{Appendix}}

%\appendix

\setcounter{section}{0} 
\renewcommand{\thesection}{S\arabic{section}}
 
\setcounter{equation}{0}  
\renewcommand{\theequation}{S\arabic{equation}}

\setcounter{figure}{0}    
\renewcommand{\thefigure}{S\arabic{figure}}

\setcounter{table}{0}    
\renewcommand{\thetable}{S\arabic{table}}

\section*{{Supporting Information (SI)}}
\section{\label{appendix:model_development}  Model Development} 
 
We follow an approach similar to that of Zhang et al. \citep{zhang2017random, SafdariJBD2022} and  calculate the probability of the existence of edges by solving a master equation. Defining $p_{ij,k}^{(s)}(t)$ as the probability of having $k$ edges of type $s$, where $s$ represents regular or anomalous connections, between nodes $i$ and $j$ at time $t$, we can express the evolution of this probability using the following master equation:

\begin{equation}
\frac{d p_{ij,k}^{(s)}(t)}{dt} = \Lambda^{(s)}_{ij}(t) \, p_{ij,k-1}^{(s)}(t) + (k+1) \Gamma^{(s)}\, p_{ij,k+1}^{(s)}(t)- \left(\Lambda^{(s)}_{ij}(t)+k\, \Gamma^{(s)} \right)p_{ij,k}^{(s)}(t),
\label{eqSI:mst1}
\end{equation}
 
where $\Lambda^{(s)}_{ij}(t)= \lambda_{ij}(t)$ and $\Gamma^{(s)}=\gamma$ stand for regular edges, ($s=r$), while $\Lambda^{(s)}_{ij}(t)= \ell$, and $\Gamma^{(s)}=\zeta$ for anomalous edges ($s=a$). To solve this equation, we use a generating function approach \citep{gardiner2004handbook} by defining $g^{(s)}(z,t)=\sum_{k=0}^{\infty} p_{ij,k}^{(s)}(t) z^{k}$. The solution for the generating function is given by \citep{Safdaridyn_2022}:

\begin{equation}\label{eqn:gzt}
g^{(s)}(z,t) = f\left[(z-1) e^{-t \Gamma^{(s)}}\right] e^{\frac{(z-1) \Lambda^{(s)}_{ij}(t)}{\Gamma^{(s)} } }, 
\end{equation}
which can be expanded in terms of $z$ to give us $p_{ij,k}^{(s)}(t)$ (more details in Sec. \ref{appendix:master_eq}). For each type of edges, anomalous or regular edge, there are  four possible transitions from time $t-1$ to $t$: 1) there is no edge neither at time $t-1$, nor at $t$; 2)  the appearance of an edge from non-edge, 3) disappearance of an existing edge, and 4) an existing edge remains; with the following probabilities, respectively,  
\bea 
&& p^r_{0\rightarrow 0} = e^{-\beta \times \lambda_{ij}(t)}, \hspace{2.5cm} p^a_{0\rightarrow 0} = e^{-\phi \times \ell} \nonumber  \\
&& p^r_{0\rightarrow 1} =    \beta \times \lambda_{ij}(t) e^{-\beta \times \lambda_{ij}(t)},\hspace{0.95cm}p^a_{0\rightarrow 1} =    \phi\times \ell e^{-\phi \times \ell}\nonumber  \\
&& p^r_{1\rightarrow 0} =   \beta \, e^{-\beta \times \lambda_{ij}(t)},\hspace{2.2cm}p^a_{1\rightarrow 0} =\phi \, e^{-\phi \times \ell}\nonumber  \\
&& p^r_{1\rightarrow 1} =    (1-\beta)  e^{-\beta \times \lambda_{ij}(t)},\hspace{1.35cm}p^a_{1\rightarrow 1} =(1-\phi)  e^{-\phi \times \ell} \qquad,
\label{eqSI:TrasitionRates}\eea
where $\beta=1- e^{-\gamma}$, and $\phi=1- e^{-\zeta}$.  Hence, we can find the transition probability for regular and anomalous edge as follows, 
\bea \label{eqSI:TransProb}
P(A_{ij}(t)|A_{ij}(t-1),\Theta) &&= \rup{P_{a}(A_{ij}(t)|A_{ij}(t-1),\Theta)}^{Z_{ij}} \times \rup{P_{r}(A_{ij}(t)|A_{ij}(t-1),\Theta)}^{1-Z_{ij}}  \nonumber  \\ 
&&  =  \rup{ \rup{p^a_{0\rightarrow 0}}^{(1-A(t-1))(1-A(t))}  \rup{p^a_{0\rightarrow 1}}^{(1-A(t-1))A(t)}    \rup{p^a_{1\rightarrow 0}}^{A(t-1)(1-A(t))}     \rup{p^a_{1\rightarrow 1}}^{A(t-1)A(t)} }^{Z_{ij}}  \nonumber  \\
&&  \times \rup{ \rup{p^r_{0\rightarrow 0}}^{(1-A(t-1))(1-A(t))}  \rup{p^r_{0\rightarrow 1}}^{(1-A(t-1))A(t)}    \rup{p^r_{1\rightarrow 0}}^{A(t-1)(1-A(t))}     \rup{p^r_{1\rightarrow 1}}^{A(t-1)A(t)} }^{1-Z_{ij}}   \nonumber  \\ 
&&  =  \rup{  e^{-\phi \times \ell} \, \rup{\phi\times \ell}^{(1-A(t-1))A(t)}    \rup{\phi }^{A(t-1)(1-A(t))}  \rup{(1-\phi) }^{A(t-1)A(t)} }^{Z_{ij}}  \nonumber  \\
&&  \times \rup{ e^{-\beta \times \lambda_{ij}(t)} \, \rup{\beta \times \lambda_{ij}(t)}^{(1-A(t-1))A(t)}    \rup{\beta}^{A(t-1)(1-A(t))}  \rup{(1-\beta) }^{A(t-1)A(t)}}^{1-Z_{ij}}\,.
\eea

Given the definition of the joint distribution in \Cref{eq:joint_dist}, this  leads to the following time-dependent, log-posterior:   
\begin{small}
\bea 
\mathcal{L}(T, \Theta)&&=\log[P(\{A(t)\}|\{A(t-1)\},\Theta )]    \nonumber \\ \nonumber \\
&&=-\sum _{Z} q(Z)\,\log q(Z)+\sum _{i,j}\sum _{Z} q(Z)\,  \left\{\log \rup{\left[ \pois(A_{ij}; \lambda_{ij}(0))\,(1-\mu)\right]^{(1-Z_{ij})} \left[ \pois(A_{ij}; \pi)\,\mu\right]^{Z_{ij}}}\right.\nonumber \\ 
&& \left. +\sum _{t=1}^T \log \rup{ [e^{-\beta \times\lambda_{ij}(t)}]^{\left(1-A_{ij}(t-1)\right) (1-A_{ij}(t)})  \times  \left[\beta  \times  \lambda_{ij}(t)\right]^{\left(1-A_{ij}(t-1)\right) A_{ij}(t)}  \times  \beta ^{A_{ij}(t-1) \left(1-A_{ij}(t)\right)}  \times(1-\beta )^{A_{ij}(t-1) A_{ij}(t)} }^{(1-Z_{ij})}  \right.  \nonumber \\ \nonumber \\
&&\left.  +\sum _{t=1}^T \log \rup{ [e^{-\phi \times\ell}]^{\left(1-A_{ij}(t-1)\right) (1-A_{ij}(t)})  \times  \left[\phi  \times  \ell\right]^{\left(1-A_{ij}(t-1)\right) A_{ij}(t)}  \times  \phi ^{A_{ij}(t-1) \left(1-A_{ij}(t)\right)}  \times(1-\phi )^{A_{ij}(t-1) A_{ij}(t)} }^{Z_{ij}}  \right\} \, . \qquad
\label{eqSI:Lconv0}\eea
\end{small}

\section{\label{appendix:master_eq} Master Equation}
  
To solve the master equation in Eq. \ref{eqSI:mst1}, we multiply both sides by $z^{k}$ and sum over $k$;  then defining the  generating function $g^{(s)}(z,t)=\sum_{k=0}^{\infty} p_{k}^{(s)}(t) z^{k}$, we get, 
 
\bea\label{eqn:mst-gn} 
\f{\partial g^{(s)}(z,t)}{\partial t} &=& \Lambda^{(s)}_{ij}(t) \, z\,g^{(s)}(z,t)+ \Gamma^{(s)}  \f{\partial g^{(s)}(z,t)}{\partial z} - \Lambda^{(s)}_{ij}(t) \,g^{(s)}(z,t)-z\, \Gamma^{(s)} \, \f{\partial g^{(s)}(z,t)}{\partial z}  \quad \nonumber \\
&& =(z-1)\left[ \Lambda^{(s)}_{ij}(t)  \,g^{(s)}(z,t)-\Gamma^{(s)}  \, \f{\partial g^{(s)}(z,t)}{\partial z}  \right] \quad.
\eea

By replacing time dependent  $\Lambda^{(s)}_{ij}(t)$ in Eq. \ref{eqn:mst-gn}, by its expected value, we reach the following relation for $g^{(s)}(z,t)$, 

\bea
\label{eqn:gzt1}
g^{(s)}(z,t) =  f\left[(z-1) e^{-t\,\Gamma^{(s)}}\right]  e^{\frac{(z-1) \Lambda^{(s)}_{ij}}{\Gamma^{(s)} } }, \quad
\eea
where $f(x)$ is any once-differentiable function of its argument satisfying $f(0) = 1$, \cite{zhang2017random,Safdaridyn_2022}.  We can assume that at $t=0$, there were no edges between two specific nodes, hence by setting $t=0$ in Eq. \ref{eqn:gzt1}, 
we find $f(x)= e^{-\f{x  }{\Gamma^{(s)}} \,  \Lambda^{(s)}_{ij}} $. For the regular edges, we assume the mean as $\frac{ \Lambda^{(s)}_{ij}}{\Gamma^{(s)} } = \lambda_{ij}$. Therefore, at $t=1$, 

\bea
\label{eqn:gzt2}
g^{(r)}(z,1) &&=   \exp \left\{(z-1) (\lambda_{ij})(1-e^{-\gamma })\right\}\nonumber  \\
&& =  \exp \left\{\beta\, (z-1) (\lambda_{ij})  \right\}  \quad.
\eea

\section{Inference}\label{appendix:sec:inference}
Our goal is, given these two mechanisms, i.e., edge community membership and anomalous edges, first to determine the values of the latent parameters, $\Theta \equiv \{ u, v, w(t),\beta, \mu, \pi,\phi,\ell \}$,  which  determine  the  relationship between the hidden label $Z_{ij}$ and the data, and then, given those values, to estimate the label $Z_{ij}$ itself.

We have the posterior: 
\be\label{eqn:joint}
P(Z,\Theta| A) = \f{P(A|Z,\Theta) P(Z|\mu) P(\Theta)P(\mu)}{P(A)} \quad.
\ee
Summing over all the possible labels we have:
\be
P(\Theta| A) = \sum_{Z}P(Z,\Theta| A) \quad,
\ee
which is the quantity that we need to maximize to extract the optimal $\Theta$.
It is more convenient to maximize its logarithm, as the two maxima coincide. We use the Jensen's inequality:
\be\label{eqn:jensen}
\log P(\Theta| A) = \log\sum_{Z}P(Z,\Theta| A) \geq  \sum_{Z} q(Z)\, \log \f{P(Z,\Theta| A)}{q(Z)} \quad,
\ee
where $q(Z)$ is a variational distribution that must sum to 1. In fact, the exact equality happens when:
\be\label{eqn:q}
q(Z) = \f{P(Z,\Theta| A)}{\sum_{Z } P(Z,\Theta| A)}\quad,
\ee
this definition is also maximizing the right-hand-side of Eq.~(\ref{eqn:jensen})  w.r.t. $q$. We then need to maximize it with respect to $\Theta$ to get the latent variables. This can be done in an iterative way using Expectation-Maximization, alternating between maximizing w.r.t. $q$ using Eq.~(\ref{eqn:q}) and then maximzing Eq.~(\ref{eqn:jensen}) w.r.t. $\Theta$.\\

Defining $Q_{ij} = \sum_{Z} q(Z)\, Z_{ij}$, the expected value of $Z_{ij}$ over the variational distribution, we  obtain:
\begin{small}
\bea \label{eqSI:Lconv0q} 
&& \mathcal{L}(T, \Theta)=\log[P(\{A(t)\}|\{A(t-1)\},\Theta )]    \nonumber \\ \nonumber \\
&&=-\sum _{Z} q(Z)\,\log q(Z)+\sum _{i,j} \rup{Q_{ij}   \log \mu + (1-Q_{ij} ) \log (1-\mu)+Q_{ij}  \rup{-\pi+A_{ij}(0) \log\, \pi} } \nonumber \\ \nonumber \\
&&  +\sum _{i,j}  \Bigg\{ (1-Q_{ij} ) \rup{-\lambda_{ij}+A_{ij}(0)\,\log \lambda_{ij}}  
 + \sum _{t=1}^T \Bigg\{  (1-Q_{ij})\rup{- \beta\, \lambda_{ij}(t)+\rup{(1-A_{ij}(t-1)) A_{ij}(t)+ A_{ij}(t-1) (1-A_{ij}(t) ) }\log \beta \right.  \nonumber \\ \nonumber \\
&&\left. + (1-A_{ij}(t-1)) A_{ij}(t)\log \lambda_{ij}(t)+A_{ij}(t-1) A_{ij}(t) \log (1-\beta) } +Q_{ij} \rup{- \phi\, \ell + (1-A_{ij}(t-1)) A_{ij}(t)\log \ell} \nonumber \\ \nonumber \\
&& + Q_{ij}\rup{\rup{ (1-A_{ij}(t-1)) A_{ij}(t)+ A_{ij}(t-1) (1-A_{ij}(t) ) } \log \phi +A_{ij}(t-1) A_{ij}(t) \log (1-\phi)}\Bigg\} \Bigg\} \quad.
\eea
\end{small}

To deal with the summation in the $\log$, we again apply Jensen's Inequality, 
\bea
&&\log \, \beta \sum _{k,q}   u_{ik} v_{jq} w_{kq}  \geq \sum _{k,q}  \rho_{ijkq}  \log \beta \, u_{ik} v_{jq} w_{kq}  - \sum _{k,q}   \rho_{ijkq}  \log  \rho_{ijkq}\quad. 
\eea
The equality  will be established when, 
\be
\rho_{ijkq} =\frac{u_{ik}v_{jq}w_{kq} }{\sum _{k,q}  u_{ik} v_{jq} w_{kq}} \quad.
\ee
It leads to the log-posterior as a function of the probability distributions, 
\begin{small}
\bea \label{eqSI:Lconv0J}  
\mathcal{L}(T, \Theta)&=&\log[P(\{A(t)\}|\{A(t-1)\},\Theta )]   =  \nonumber \\ \nonumber \\
&&-\sum _{Z} q(Z)\,\log q(Z)+\sum _{i,j} \rup{Q_{ij}  \log \mu + (1-Q_{ij}) \log (1-\mu)}   \nonumber \\
&& +\sum _{i,j}  \Bigg\{ (1-Q_{ij}) \rup{-\sum _{k,q}  u^{(0)}_{ik} v^{(0)}_{jq} w_{kq}(0)+A_{ij}(0)\,(\sum _{k,q}\rho_{ijkq}\, \log u^{(0)}_{ik}v^{(0)}_{jq}w_{kq}(0)-\sum _{k,q}\rho^{(0)}_{ijkq}\, \log \rho^{(0)}_{ijkq} )} \nonumber \\ \nonumber \\
&& + Q_{ij}  \rup{-\pi+A_{ij}(0) \log\, \pi}  \nonumber \\ \nonumber \\
&& + \sum _{t=1}^T \Bigg\{  (1-Q_{ij} )\rup{- \beta\, \sum _{k,q}  u_{ik} v_{jq} w_{kq}(t)  + (1-A_{ij}(t-1)) A_{ij}(t)(\log \beta+\sum _{k,q}\rho_{ijkq}\, \log u_{ik}v_{jq}w_{kq}(t)-\sum _{k,q}\rho_{ijkq}\, \log \rho_{ijkq}) \right.  \nonumber \\ \nonumber \\
&&\left.  + A_{ij}(t-1) (1-A_{ij}(t) ) \log \beta +A_{ij}(t-1) A_{ij}(t) \log (1-\beta)} + Q_{ij} \rup{- \phi\, \ell + (1-A_{ij}(t-1)) A_{ij}(t)\log \ell}\nonumber \\ \nonumber \\
&& + Q_{ij}\rup{\rup{ (1-A_{ij}(t-1)) A_{ij}(t)+ A_{ij}(t-1) (1-A_{ij}(t) ) } \log \phi  +A_{ij}(t-1) A_{ij}(t) \log (1-\phi)}\Bigg\} \Bigg\}\quad.
\eea  
\end{small}

We initiate the process by taking the derivative of Eq.~(\ref{eqSI:Lconv0J}), concerning the individual parameters, such as $u_{ik}$. We make the assumption of a uniform prior with respect to $\Theta$, but we can readily integrate more intricate selections if required. Furthermore, we examine static values for $u$ and $v$, while considering dynamic variations for $w$. We define $\hat{A}_{ij}(t) = A_{ij}(t) (1-A_{ij}(t-1))$ when $t>0$, $\hat{A}_{ij}(0)=A_{ij}(0)$, and $\hat{\beta}(t)=1$ for $t=0$, and $\hat{\beta}(t)=\beta$ for $t>0$.
\be
 \f{\partial \mathcal{L}(T, \Theta)}{\partial u_{ik}}  = 0  \quad, 
\ee 
\begin{footnotesize}
\bea
&& \sum_{j}\Bigg\{ [1-Q_{ij}]\left[-\sum_{q}v_{jq}w_{kq} +\sum_{q}\rho_{ijkq}\hat{A}_{ij}(0)\f{1}{u_{ij}}\right]  + \sum _{t=0}^T \left( [1-Q_{ij}] \left[-\beta\, \sum_{q}v_{jq}w_{kq} +\hat{A}_{ij}(t) \times  \sum_{q}\rho_{ijkq}\f{1}{u_{ij}}  \right]  \right) \Bigg\}=0  \quad. \nonumber \\
\eea
\end{footnotesize} 
This yields: 
\be \label{appendix:eqn:u}
u_{ik} = \f{\sum _{j,q,t}\, [1-Q_{ij}]\, \rho_{ijkq}(t)\,\hat{A}_{ij}(t) }{\sum _{j,q}   v _{jq}\, [1-Q_{ij}]\sum_{t=0}^T  \hat{\beta}(t) \,w_{kq}(t)} \quad. 
\ee

We find similar expression for $v_{ik}$ and $w_{kq}$.
\be \label{appendix:eqn:v}
v_{ik} = \f{\sum _{j,q,t}\,  [1-Q_{ij}]\, \rho_{ijkq}(t)\,\hat{A}_{ji}(t) }{\sum _{j,q}   u_{jq} \,[1-Q_{ij}]\sum_{t=0}^T  \hat{\beta}(t) \,w_{kq}(t)} \quad, 
\ee

\be\label{appendix:eqn:w} 
w_{kq}(t) = \f{\sum_{i,j}\, [1-Q_{ij}]\,\rho_{ijkq}\,\hat{A}_{ij}(t)}{\sum _{i,j}\, [1-Q_{ij}]\hat{\beta}(t) \, u_{ik}\,v_{jq}} \quad.
\ee 

Following the same procedure, we find the expressions for $\mu$  and $\ell$: 

\be \label{appendix:eqn:mu}
\mu = \f{1}{N(N-1)/2}\sum_{i<j} Q_{ij} \quad,
\ee

\be \label{appendix:eqn:pi}
\pi = \f{\sum_{i,j} \, Q_{ij}\, \hat{A}_{ij}(0)}{\sum_{i,j}  Q_{ij}}\quad,
\ee

\be \label{appendix:eqn:ell}
\ell = \f{\sum_{i,j}\,\sum_{t=1}^{T}\,Q_{ij}\, \hat{A}_{ij}(t)}{T\,\sum_{i,j} \, \phi\, Q_{ij}}\quad.
\ee

The static parameters $\beta$ and $\phi$ lack closed-form updates. Hence, to determine their values, we need to employ numerical root-finding methods on their respective equations: 
\begin{align}\label{appendix:eqn:beta}
\sum_{i,j}\,\sum_{t=1}^{T}\, [1-Q_{ij}] \Bigg\{-\beta \rup{\lambda_{ij}(t)  +\f{1}{1-\beta} A_{ij}(t-1) A_{ij}(t) }
+\rup{\hat{A}_{ij}(t)+A_{ij}(t-1)(1-A_{ij}(t))}\Bigg\}=0 \, \quad,
\end{align} 
and
\begin{align}\label{appendix:eqn:phi}
\sum_{i,j}\sum_{t=1}^{T}\, Q_{ij}\, \Bigg\{-\phi \rup{\ell +\f{1}{1-\phi} A_{ij}(t-1) A_{ij}(t) }
+\rup{\hat{A}(t)+A_{ij}(t-1)(1-A_{ij}(t))}\Bigg\}=0 \, \quad.
\end{align}

To estimate $q(Z)$, we insert the estimated parameters into \cref{eqn:joint}:
\bea
q(Z) &=& \f{\prod_{i,j}  \rup{\pois(A_{ij}(0);\pi) F_a(\phi, \ell, A_{ij}, T) }^{Z_{ij}} \, \rup{\pois(A_{ij}(0); \lambda_{ij}(0))\, F_r(\beta, \lambda_{ij}(t), A_{ij}, T) }^{1-Z_{ij}} \,\prod_{i<j} \mu^{Z_{ij}}\, (1-\mu)^{(1-Z_{ij})}}
{\sum_{Z}\prod_{i,j}  \rup{\pois(A_{ij}(0);\pi) F_a(\phi, \ell, A_{ij}, T) }^{Z_{ij}}\, \rup{\pois(A_{ij}(0); \lambda_{ij}(0))\,F_r(\beta, \lambda_{ij}(t), A_{ij}, T) }^{1-Z_{ij}} \,\prod_{i<j} \mu^{Z_{ij}}\, (1-\mu)^{(1-Z_{ij})}} \nonumber\\  \nonumber\\
&=& \prod_{i<j} \f{ \rup{\pois(A_{ij}(0);\pi) \pois(A_{ji}(0);\pi) \, F_a(\phi, \ell, A_{ij}, T) F_a(\phi, \ell, A_{ji}, T) \, \mu}^{Z_{ij}} }
{\sum_{Z_{ij}=0,1}  \rup{\pois(A_{ij}(0);\pi) \pois(A_{ji}(0);\pi)\,F_a(\phi, \ell, A_{ij}, T) F_a(\phi, \ell, A_{ji}, T) \, \mu}^{Z_{ij}}}\qquad\\ \nonumber \\
&\times & \prod_{i<j} \f{ \rup{\pois(A_{ij}(0);\lambda_{ij}(0))\,\pois(A_{ji}(0);\lambda_{ji}(0))\, F_r(\beta, \lambda_{ij}(t), A_{ij}, T)\, F_r(\beta, \lambda_{ji}(t), A_{ji}, T)\, (1-\mu)}^{(1-Z_{ij})}}
{\sum_{Z_{ij}=0,1}  \rup{\pois(A_{ij}(0);\lambda_{ij}(0))\,\pois(A_{ji}(0);\lambda_{ji}(0))\, F_r(\beta, \lambda_{ij}(t), A_{ij}, T)\, F_r(\beta, \lambda_{ji}(t), A_{ji}, T)\,(1-\mu)}^{(1-Z_{ij})}}\qquad\\ \nonumber \\
&=& \prod_{i<j}\, Q_{ij}^{Z_{ij}}\, (1-Q_{ij})^{(1-Z_{ij})} \quad, \label{appendix:eqn:qQij} 
\eea

where
\begin{scriptsize}
\bea  \label{appendix:eqn:Qij}
&&Q_{ij} =\nonumber\\  \nonumber\\
&&\f{\mu \, \pois(A_{ij}(0);\pi) \pois(A_{ji}(0);\pi)\times F_a(\phi, \ell, A_{ij}, T)\, F_a(\phi, \ell, A_{ji}, T)}{ \mu\, \pois(A_{ij}(0);\pi) \pois(A_{ji}(0);\pi) F_a(\phi, \ell, A_{ij}, T) F_a(\phi, \ell, A_{ji}, T)+(1-\mu)\pois(A_{ij}(0); \lambda_{ij}(0))\pois(A_{ji}(0); \lambda_{ji}(0)) F_r(\beta, \lambda_{ij}(t), A_{ij}, T) F_r(\beta, \lambda_{ji}(t), A_{ji}, T) },  \nonumber \\ 
\eea
\end{scriptsize}
and 
\bea \label{appendix:eqn:F-funcs} 
&& F_r(\beta, \lambda_{ij}, A_{ij}, T) =  \prod _{t=1}^T\,\e^{-\beta\times\lambda_{ij}(t)}\rup{[\beta\,\lambda_{ij}(t)]^{(1-A_{ij}(t-1))A_{ij}(t)}[\beta]^{A_{ij}(t-1)(1-A_{ij}(t))} [1-\beta]^{A_{ij}(t-1)A_{ij}(t)} } \\
&& F_a(\phi, \ell, A_{ij}, T)  =  \prod _{t=1}^T\,\e^{-\phi\times\ell}\rup{[\phi\,\ell]^{(1-A_{ij}(t-1))A_{ij}(t)}\phi^{A_{ij}(t-1)(1-A_{ij}(t))} [1-\phi]^{A_{ij}(t-1)A_{ij}(t)} } .
\eea

\section{Covergence criteria}
The EM algorithm involves the random initialization of $\Theta \equiv \{ u, v, w(t),\beta, \mu, \pi,\phi,\ell \}$, followed by iterations over Eqs.~\ref{appendix:eqn:u}-\ref{appendix:eqn:ell}, and \ref{appendix:eqn:Qij}, until achieving convergence of the log-posterior in Eq.~(\ref{eqSI:Lconv0q}).   For the computation of $\log q(Z)$, we employ Eq.~(\ref{appendix:eqn:qQij}), representing a Bernoulli distribution.

\section{\label{appendix:sec:GM} Generative model}

Being generative, our model can be used to generate synthetic networks that include both  anomalous edges and community structure over time. To this end, we sample the parameters $\Theta \equiv \{ u, v, w(t),\beta, \mu, \pi,\phi,\ell \}$ and then, given these latent variables, we sample $Z$. Finally,  given the $Z$ and the latent variables, we can sample the adjacency matrix $A$.

To build the adjacency matrix at the first time step, i.e., $T=0$, and for a given set of community parameters as the input \cite{de2017community,Safdari2022ACD}, we sample anomalous edges from a Poisson distribution as in \Cref{eqn:A0}, with a  Bernoulli prior  as in \Cref{eqn:Zprior}. The mean value of the Poisson distribution, $\pi$,  is constant for all edges, however, its value can be chosen in order to control the ratio $\rho_{a}$ of edges being anomalous over the total number of edges. The average number of anomalous  and non-anomalous edges are $N\mu\, (1-e^{-\pi})$, and $(1-\mu)\, \sum_{i,j}(1-e^{-\lambda_{ij}(0) })$, respectively. Assuming a desired total number of edges $E$, we can  multiply $\pi, \mu$ and $\lambda_{ij}(0) $ by suitable sparsity constants that tune: i) the ratio $$\rho_{a}=\f{N\mu\, (1-e^{-\pi})}{N\mu\, (1-e^{-\pi})+(1-\mu)\, \sum_{i,j}(1-e^{-\lambda_{ij}(0) })} \quad \in \rup{0,1};$$ ii) the success rate of anomalous edges $\pi$. Once these two quantities  are fixed, the remaining sparsity parameter for the matrix $\lambda$, is estimated as:

\begin{equation}\label{appendix:eqn:}
E\,(1-\rho_{a}) = (1-\mu)\, \sum_{i,j}(1-e^{-c\lambda_{ij}(0) })\quad, 
\end{equation} 

which can be solved numerically with root-finding methods.

To construct adjacency matrix at the next time steps, we use the transition rates from \Cref{eqSI:TrasitionRates}.

\section{Results on synthetic datasets}\label{appendix:sec:res_syn} 

\paragraph{Experimental settings for synthetic networks}
We consider networks with hard communities of equal size and assortative structure. We vary the anomaly density within the range $\rho_a \in [0,1]$ and set the disappearance rate of anomalous and regular edges to $\phi=0.2$ and $\beta=0.2$, respectively.
For comparison, we establish a baseline model, termed \dmcd,  which simplifies our model to standard community detection for dynamic networks without any anomaly detection mechanism. This is obtained by setting $\mu =0$ and $\lim {\ell \to 0}$, which are kept fixed as hyper-parameters in  the inference task.
Having fixed the parameters, we generate 10 samples of networks for each value of $\rho_a$. For each network, we generate an initial state followed by up to $T$ further snapshots. The initial state is formed based on the the community structure  and anomaly density outlined in Eq.~(\ref{eqn:A0}). The successive snapshots are generated according to transition rates in Eq. (\ref{eqSI:TrasitionRates}), as explained in \cref{appendix:sec:GM}.\\
 
\paragraph{Community detection} As the presented algorithm hinges on incorporating the community structure as the regular model for tie formation between nodes,  an important aspect is to demonstrate the competence of the model in accurately retrieving the community structure.  
The results presented in \Cref{figSI:CS_AUCsynth}(a) demonstrate the effectiveness of our proposed anomaly-community detection algorithm, \dmacd, compared to the model without an anomaly detection mechanism, \dmcd. 
The plot shows the cosine similarity (CS) between the ground truth and inferred communities for synthetic networks with $N = 300$  nodes, average degree $\langle k \rangle=8$, and $K = 8$ communities, and total number of time stamps $T = 20$. 

The results indicate that \dmcd \text{} performs similar to \dmacd\  for small values of anomaly density ($\rho_a$). Nevertheless, as  the number of edges without community structure increases, the algorithm's ability in community detection decreases significantly, aligning with expectations.
On the other hand, \dmacd, which includes an anomaly detection mechanism, is greatly impacted by the number of time steps. Despite its weaker performance in community detection during the initial time steps, it exhibits continuous improvement over time.  This demonstrates the effectiveness of the proposed anomaly detection mechanism in improving the overall community detection accuracy.

\paragraph{Link prediction performance} 
In addition, we assess  the performance of \dmacd\ in the link prediction task, comparing it with \dmcd. 
 For each time step $t\in [1,T]$, we hide an individual snapshot $A(t)$ (test set), and fit the data using the previous snapshots $A(0),\dots,A(t-1)$ (training set). We test whether the model is able to predict the network's evolution using the area under the curve (AUC) calculated on the test set, which gauges the likelihood that a randomly chosen edge possesses a higher expected value than a randomly selected non-existent edge. A perfect reconstruction is indicated by a value of $1$, while $0.5$ denotes random chance.
To compute the expected value of a regular edge, Eq. (\ref{eqn:ExpAij}) is employed:
\bea\label{eqn:ExpAij} 
\Exp\rup{ A_{ij}(t)}  &=& \begin{cases}\f{ \beta\,\lambda_{ij}(t)}{1+ \beta\,\lambda_{ij}(t)} &\text{if} \,\, A_{ij}(t-1)=0\\
						 1-\beta &\text{if} \,\,  A_{ij}(t-1)=1 
						 \end{cases}	 \,.
\eea 

Similarly, the expected value of an anomalous edge is computed using Eq. (\ref{eqn:ExpAij_anomalous}):
\bea\label{eqn:ExpAij_anomalous}
\Exp\rup{ A_{ij}(t)}  &=& \begin{cases} \f{\phi\,\ell_{ij}}{1+ \phi\,\ell_{ij}} &\text{if} \,\, A_{ij}(t-1)=0\\
						 1-\phi &\text{if} \,\,  A_{ij}(t-1)=1 
						 \end{cases}	 \,.
\eea 

It is important to note that while the expected value at time $t$ relies only on the network at the preceding time step, all parameters are inferred using the entire network history (excluding the final time step, i.e. test set). Therefore, our model is trained with the complete set of snapshots ${A(0),\dots, A(t-1)}$. 
In general, both \dmacd\ and \dmcd \ demonstrate reasonable performance in link prediction. However, \dmcd \ reaches a plateau in its performance over time. In contrast, \dmacd \ exhibits a clear and consistent increase in the estimated AUC over time, as illustrated in \Cref{figSI:CS_AUCsynth}(b). Notably, while the estimated AUC values by \dmacd \ converge to a maximum over time across all ranges of anomaly density, $\rho_a$, \dmcd \ exhibits lower AUC values, particularly for higher values of $\rho_a$.  In other words, as the number of anomalous edges increases, \dmcd \ loses its effectiveness in link prediction. 

\paragraph{Estimating parameters}

Considering that \dmacd\ is a generative model, it has  the capability to learn the model parameters, $\Theta \equiv \{ u, v, w(t),\beta, \mu, \pi,\phi,\ell \}$, from the provided dataset. These parameters enable the algorithm to generate synthetic datasets with statistical features mirroring those of the original dataset. Therefore, accurate inference of the model parameters is crucial for the model's proficiency. 
\Cref{tabSI:recoveredVAR} presents the estimated values of the latent variables in synthetic data as we vary the density of anomalous edges $\rho_{a}$, illustrating the ability of the algorithm in retrieving values similar to the ground truth value.

%----------------------------------------------------------------------------
 \begin{figure}[t]  
 \includegraphics[width=1\linewidth]{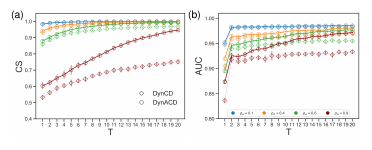}    
  \caption{\textbf{Community detection and Link Prediction in synthetic networks.}   a)  Cosine similarity (CS) between ground truth and inferred communities. The model without anomaly detection mechanism, \dmcd, performs similar to \dmacd\,  for small values of anomaly density, $\rho_a$. However, by increasing $\rho_a$, i.e., more edges without community  structure, the algorithm's ability in community detection decreases, as it is expected. b)  \dmacd  \text{} exhibits a clear and consistent increase in the estimated (AUC) over time.   Synthetic networks were generated  with $N = 300$ nodes, average degree $\langle k \rangle=8$, and $K = 8$. In addition the model parameters were fixed to $\beta=0.2$, $\ell=0.2$, $\phi=0.2$.  Lines are averages and standard deviations over $10$ sampled networks. Model with Anomaly mechanism: True ($\circ$), and False ($\diamond$). \\}  \label{figSI:CS_AUCsynth}
\end{figure}
%----------------------------------------------------------------------------

% --------------------------------------- Table synthetic data recovered parameters  --------------------------------------------------------
\begin{table}[h]
\caption{{\bf {Estimated latent variables}} Synthetic networks with $N = 300$ nodes, average degree $\langle k \rangle=8$, and $K = 8$ communities of equal-size unmixed group membership generated with our generative model. Ground truth values are set to: $\beta=0.2$, $\ell=0.2$, $\phi=0.2$.  We report estimated averages and standard deviations over 10  samples for each of the main model's parameters.  }
\begin{center}
\begin{ruledtabular}  
\begin{tabular}{ccccc}
$\rho_{a}$  & $\pi$ & $\phi$ & $\ell$ & $\beta$ \\ \hline
0.1 &    		$0.2434 \pm 0.0220$&           $0.2091 \pm 0.0078$   & $0.2151  \pm  0.0094$&           $0.1994 \pm 0.0019$\\
0.4 & 			$0.2360 \pm 0.0117$&           $0.2085 \pm 0.0043$   & $0.2072  \pm  0.0057$&           $0.1981 \pm 0.0033$\\
0.6 &           $0.2285 \pm 0.0070$&           $0.2096 \pm 0.0054$   & $0.2044  \pm  0.0079$&           $0.1963 \pm 0.0035$\\
0.9 &           $0.2249 \pm 0.0088$&           $0.2112 \pm 0.0025$   & $0.1966  \pm  0.0046$&           $0.1904 \pm 0.0043$\\
\end{tabular} 
\end{ruledtabular} 
\end{center}
\label{tabSI:recoveredVAR}
\end{table}

\section{\label{appendix:data_desc} REAL--WORLD DATA SETS: DATASET DESCRIPTION}

\Cref{tabSI:data_desc} provides a summary of the key characteristics of the studied datasets. The datasets  have undergone a preprocessing step that involved the splitting the edges in time snapshots,  the removal of self-loops, retaining only nodes with both non-zero incoming and outgoing edges, and keeping only the giant connected components. In the case of  citation network (here: SCC2016 ), it requires an additional pre-process of extracting a network of author-author from a network of paper-citation; hence  an edge means that an author cites another author. 

% --------------------------------------- Table real data description  --------------------------------------------------------------------------
\begin{table}[h]  
\caption{{\bf {Real-world datasets description.}}} 
\begin{ruledtabular}
\begin{tabular}{lllllll}
 \textbf{Network} & \textbf{Abbreviation}    &$N$ &$E$& $T$  & \textbf{Ref.}\\ \hline
Data on European football clubs' player transfers \hspace{10pt}  & Transfermarkt   \hspace{10pt}  & $261$   \hspace{10pt}     & $7613$     \hspace{10pt}   & $5$ \hspace{10pt}& \cite{transfersmarkt_ewenme}    \\   
Network of United States Air Transportation   \hspace{10pt}  & US-Air  \hspace{10pt}  & $867$   \hspace{10pt}     & $8039$     \hspace{10pt}  & $6$     \hspace{10pt}& \citep{Charyyevusairpot2020}    \\
Email Eu core network   &\eum\text{}         & $891$     & $18679$    &$12$        & \cite{Leskovec2007}    \\  
 Statistics Citation & SCC2016        & $1159$   & $4504$     &$4$  &    \citep{Ji2016}    \\ 
The UC Irvine messages network   \hspace{10pt}  & UCI Messages \hspace{10pt}  & $1810$   \hspace{10pt}     & $18693$      \hspace{10pt}  & $6$     \hspace{10pt}& \citep{konect:2017:opsahl-ucsocial}    \\
\end{tabular}
\end{ruledtabular} 
\label{tabSI:data_desc}
\end{table}

\subsubsection{US Air Transport Networks}
The dataset used in this study provides the dynamics of the United States air transportation network from 2007 to 2016.  
The original format of the dataset contains three different representations of nodes (airport, city, state) and edges (number of passengers, number of flights, amount of freight), resulting in a total of nine different networks. The dataset allows for a comprehensive analysis of the air transportation system, considering various perspectives and levels of granularity \citep{Charyyevusairpot2020}.
For this study, the we focused on the period from 2011 to 2016 and used airports as nodes. The edges represent the flights between airports, with the weight of the edges indicating the number of flights. This US Air Transport Networks contribute to a detailed understanding of the connectivity and activity within the air transportation network during the specified time frame.\\

\subsubsection{Transfermarkt}
The Transfermarkt dataset contains information on player transfers among eight European men football clubs starting from the 1992/93 season \cite{transfersmarkt_ewenme} . Transfermarkt is a dedicated website that offers a wide range of data on football transfers, market values, rumors, and statistics. The dataset is structured into separate files for different leagues, including the English Premier League, French Ligue 1, German 1.Bundesliga, Italian Serie A, Spanish La Liga, Portuguese Liga NOS, Dutch Eredivisie, and Russian Premier Liga. Key variables within the dataset include club name, player name, position, transfer fee, transfer movement, transfer period, league name, year, season, and country. Our analysis focuses on the dataset spanning the 2008/2009 season to the 2022/2023 season, and for algorithmic application, we segmented the dataset into five time snapshots, each spanning a four-year duration. Here,  edges indicate transfers of at least one player between two clubs in a given time window. 
 
\subsubsection{UCI Messages}
The UC Irvine messages network dataset is a directed network that contains sent messages between users in an online community of students from the University of California, Irvine \citep{konect:2017:opsahl-ucsocial}. The dataset consists of nodes that represent the users and directed edges that represent the sent messages.\\

\subsubsection{EU email network} 
The Email-Eu-core network is based on internal emails exchanged among members of a large European research institution. At each time step, a directed edge from node $i$ to node $j$ is created if $i$ sent an email to $j$. The dataset spans a period of 803 days. For our analysis, we examined the dataset dynamics by dividing it into monthly intervals. Specifically, we partitioned the edges into monthly segments (\eum) and selected snapshots from the first recorded year.

\subsubsection{Statistics Citation dataset} 
The Citation network for statisticians is derived from research papers published in four top statistics journals from $2003$ to the first half of $2012$. We construct the network by sampling data from $2003$ to $2007$ and dividing it into annual intervals. This results in a citation network spanning four years, where nodes represent authors and a directed edge from node $i$ to node $j$ at time step $T$ indicates that $i$ cited $j$'s papers in that year.

\section{Results on REAL-WORLD DATA SETS}\label{appendix:sec:res_RD}  
In this section, we present a detailed examination of the results obtained from anomaly detection experiments conducted on real-world datasets. Specifically, we analyze the modified Transfermarkt dataset with injected anomalous edges at a density of $\rho_a=0.1$.  The histogram in \Cref{figSI:onesample_injection}(a)  provides a visual representation of the distribution of inferred anomaly scores ($Q$) assigned to edges by  \dmacd. It effectively discerns between true labels, emphasizing the density patterns of scores for both ``Anomalous'' and ``Regular'' instances. It is evident that edges labeled as ``Regular'' exhibit an inferred anomaly score ($Q$) with a peak around zero. In contrast, the inferred values of $Q$ for injected edges—those labeled as ``Anomalous''—extend across higher values, trending towards one. 
The confusion matrix in \Cref{figSI:onesample_injection}(b) also visualizes the classification performance of the model, particularly in distinguishing between anomalous and regular edges. \\

% --------------------------------------- Figure Real World Datasets----------------------------------------------------- 
\begin{figure}[t]  \label{figSI:onesample_injection}
\includegraphics[width=0.9\linewidth]{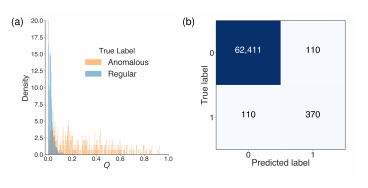}      
\caption{\textbf{Anomaly Detection in the Transfermarkt dataset with injected anomalous edges.}  a) Anomaly Score Distribution ($Q$) Separated by True Labels. The histogram distinguishes between true labels, highlighting the density of scores for both ``Anomalous'' and ``Regular'' instances. b) The confusion matrix illustrates the classification performance of the model, breaking down the predicted and true labels for each class. The diagonal elements represent correctly classified instances, while off-diagonal elements indicate misclassifications. We achieve a recall rate of $0.77$ and an AUC(Z) of $0.96$ on the Transfermarkt dataset. Here, $\rho_a=0.1$. } 
\end{figure}    
% ---------------------------------------

The results inferred by \dmacd{} and presented in \Cref{tab:UdineseWatford_lambdaQ} align well with our expectations from the algorithm. Notably, we observe detailed specificity in the transfer information between Udinese Calcio and Watford FC, along with insights into the transfers involving NEC Nijmegen and CA Osasuna.
In both cases, the expected value of observing a regular transfer between the two involved clubs is negligible, i.e., $\lambda_{ij} \rightarrow 0$. Therefore, it is reasonable to infer the existence of an irregular interaction between such clubs, i.e., $Q_{ij} = 1$. This observation highlights that not only can the consistency in the appearance and disappearance of transfers over time contribute to a higher value for $Q_{ij}$, as seen in the case of Udinese Calcio and Watford FC, but also a single-time transfer out of community membership can yield the same result, as the transfer between NEC Nijmegen and CA Osasuna. \\

% --------------------------------------- Table Udinese_Calcio  ---------------------------------------------------- 
\begin{table}[b]
\caption{\textbf{Transfermarkt Dataset:} Instances of transfers between clubs exhibiting a low expected value for tie formation based on community membership. Consequently, these interactions are identified as irregular, regardless of whether they feature high or low appearance rates.}
\begin{center}
\begin{ruledtabular}  
\begin{tabular}{llcccccccc}
Source Club ($i$)&      Target  Club ($j$)&  $A_{ij}(0)$ &  $A_{ij}(1)$ &  $A_{ij}(2)$ &  $A_{ij}(3)$ &  $A_{ij}(4)$ &  $\sum_{t}A_{ij}(t)$ &  Q$_{ij}$ &  $\langle \lambda_{ij}^{t}\rangle$\\ \hline \\
 Udinese Calcio &     Watford FC &          0 &          1 &          1 &          1 &          1 &      4 & 1.0 &      0.049 \\
    Watford FC & Udinese Calcio &          0 &          0 &          1 &          1 &          1 &      3 & 1.0 &      0.013 \\
    \hline \\
NEC Nijmegen &   CA Osasuna &          1 &          0 &          0 &          0 &          0 &      1 & 1.0 &      0.000 \\
  CA Osasuna & NEC Nijmegen &          0 &          1 &          0 &          0 &          0 &      1 & 1.0 &      0.000 \\
\end{tabular} 
\end{ruledtabular} 
\end{center}
\label{tab:UdineseWatford_lambdaQ}
\end{table}
% ---------------------------------------  
 
During our experiments, in the data preprocessing, we  aggregated the transfer data of two Italian clubs, Inter Milan and AC Milan. Surprisingly, the algorithm inferred a very high value for $Q_{ij}$ for many of  the interactions of the aggregated club, referred to as Milan. This observation stemmed from a higher rate of connections between the aggregated club and numerous others, showcasing \dmacd{}'s ability to detect irregularities. Examples of such inferred $Q_{ij}$s are provided in \Cref{tabSI:Milan_InterMilan}.\\

% --------------------------------------- Table Udinese_Calcio  ---------------------------------------------------- 
\begin{table}[t]
\caption{\textbf{Transfermarkt Dataset with mislabeled nodes. } Instances of transfers flagged as anomalous in a manipulated dataset where Inter Milan and AC Milan are merged into a unique node ``Milan'' due to mislabeling in the data collection process. Several transfers involving ``Milan'' are flagged as anomalous, including transfers with other teams with compatible communities, as shown by a releatively high value of the average $\langle \lambda_{ij}^{t}\rangle=\sum_{t}\lambda_{ij}^{t}/T$.}
\begin{center}
\begin{ruledtabular}  
\begin{tabular}{llcccccccc}
Source Club ($i$)&      Target  Club  ($j$)&  $A_{ij}(0)$ &  $A_{ij}(1)$ &  $A_{ij}(2)$ &  $A_{ij}(3)$ &  $A_{ij}(4)$ &  $\sum_{t}A_{ij}(t)$ &  Q$_{ij}$ &  $\langle \lambda_{ij}^{t}\rangle$\\ \hline \\
 Milan & 	Genoa &          1 &          1 &          1 &          1 &          1 &      5 & 0.996 &      0.300\\
    Genoa&	Milan&          1 &          1 &          1 &          1 &          1 &      5 & 0.996 &      0.239 \\
Milan&	UC Sampdoria	 &               1 &          1 &          1 &          1 &          1 &      5& 0.966 &      0.197 \\
 UC Sampdoria	& Milan &          1 &          1 &         1 &         1 &          0 &     4 & 0.966 &     0.223 \\
 AS Roma	& Milan &          1 &          1 &         1 &         1 &          1 &     5 & 0.945 &     0.238 \\
  Milan & AS Roma&          1 &          1 &         1 &         1 &          0 &     4 & 0.945 &     0.320 \\
 Parma	& Milan &          1 &          1 &         1 &         1 &          0 &     4 & 0.901 &     0.181 \\
  Milan & Parma&          1 &          1 &         1 &         1 &          1 &     5 & 0.901 &     0.133 \\
 Manchester City	& Milan &          1 &          1 &         1 &         1 &          1 &     5 & 0.658 &     0.150 \\
  Milan & Manchester City&          1 &          1 &         0 &         0 &          0 &     2 & 0.658 &     0.176 \\
\end{tabular} 
\end{ruledtabular} 
\end{center}
\label{tabSI:Milan_InterMilan}
\end{table}
% ---------------------------------------

\Cref{figSI:Transfermarket_community_u} highlights the same points as mentioned in \Cref{sec:res:Transfermarkt}. However, this plot illustrates  the out-going (soft) community membership of clubs.

\section{Embedding methods from deep learning}\label{appendix:sec:embeddings}  
In machine learning several approaches have been developed for anomaly detection in networks \cite{YuNetWalk2018,Ma2023Akoglu}. Many of them rely on extra information on nodes, e.g., node attributes or features, which is outside the scope of the problem presented in our work. In scenarios where only the network structure is given as input, as is the case here, a common approach is to utilize embeddings. Similar to our memberships $u$ and $v$, embeddings are vectors assigned to nodes. Like our model, these embeddings are then utilized to detect anomalous edges. The various approaches differ in the way these embeddings are learned. For instance, the Netwalk algorithm \citep{YuNetWalk2018} employs techniques from deep learning.\\
The main difference between these approaches and the one considered in this paper is that our model postulates explicitly the presence of anomalous edges. This is because in our case, the probability of an edge is explicitly defined in terms of $u$ and $v$ (and the other parameters). As a consequence, we can obtain automatically (without further post-processing) and unambiguously (with the precise definition of $Z_{ij}$ in terms of the latent variables) the estimate of this probability. In contrast, embedding methods do not formulate explicitely this probability. Instead, they make a broader assumption that embeddings explain the network structure, hence, one can use them a posteriori to determine what is an anomalous and what is a regular edge. A main consequence of this broader assumption  is that one has to make arbitrary decisions in the post-processing to determine how the embeddings should be used to estimate the probability of an edge being anomalous. \\
One important arbitrary decision is how to transform a representation that is assigned to nodes (the embedding vectors $u_{i}$ and $v_{j}$) into a representation on edges $(i,j)$. For instance, Netwalk considers various functions to combine the node representations, with best results obtained with the Hadamard product $ z_{ij} = u_{i} \odot v_{j}$ \citep{Ma2023Akoglu}. In our case instead, we have automatically the estimated $Q_{ij}$ as a function of the inferred $u$ and $v$, a quantity that has a clear interpretation as the expected value of the posterior of $Z_{ij}$.\\
In addition, once the edge representation $z_{ij}$ is selected, one has to further decide how to use its inferred value to assign the probability for an edge to be anomalous. A standard approach is to run the k-means clustering algorithm on the edge representation vectors to cluster edges in a space (not to be confused with the node embedding space). Then, one computes the distance between each edge and its cluster centre and edges further away from the cluster are considered anomalous, with the distance used as a score to quantify the magnitude of being anomalous. All these steps though require further arbitrary decisions and parameters to be tuned, e.g. what clustering algorithm, the number of clusters in this new edge representation space, the exact mapping between the distance from their clusters or other clusters and the anomaly score. Further complexities are added in case of dynamic networks, where embeddings may change in time.\\
In the context of our work, we applied the Netwalk algorithm to real-world datasets studied. However, the performance of the Netwalk algorithm was found to be very poor when compared to \dmacd. It is difficult to pinpoint exactly the cause of this, as any of the decisions above could impact performance. Consequently, while we provide a comprehensive discussion on the reasons behind these difference in performance, we refrain from detailing the specific results obtained from the application of the Netwalk algorithm to the datasets studied. More investigations would be needed to understand how to calibrate these models for a fair comparison, this is beyond the scope of our work.\\

% --------------------------------------- Figure clubs_community_u 
\begin{figure}[htbp] 
\includegraphics[width=1. \linewidth]{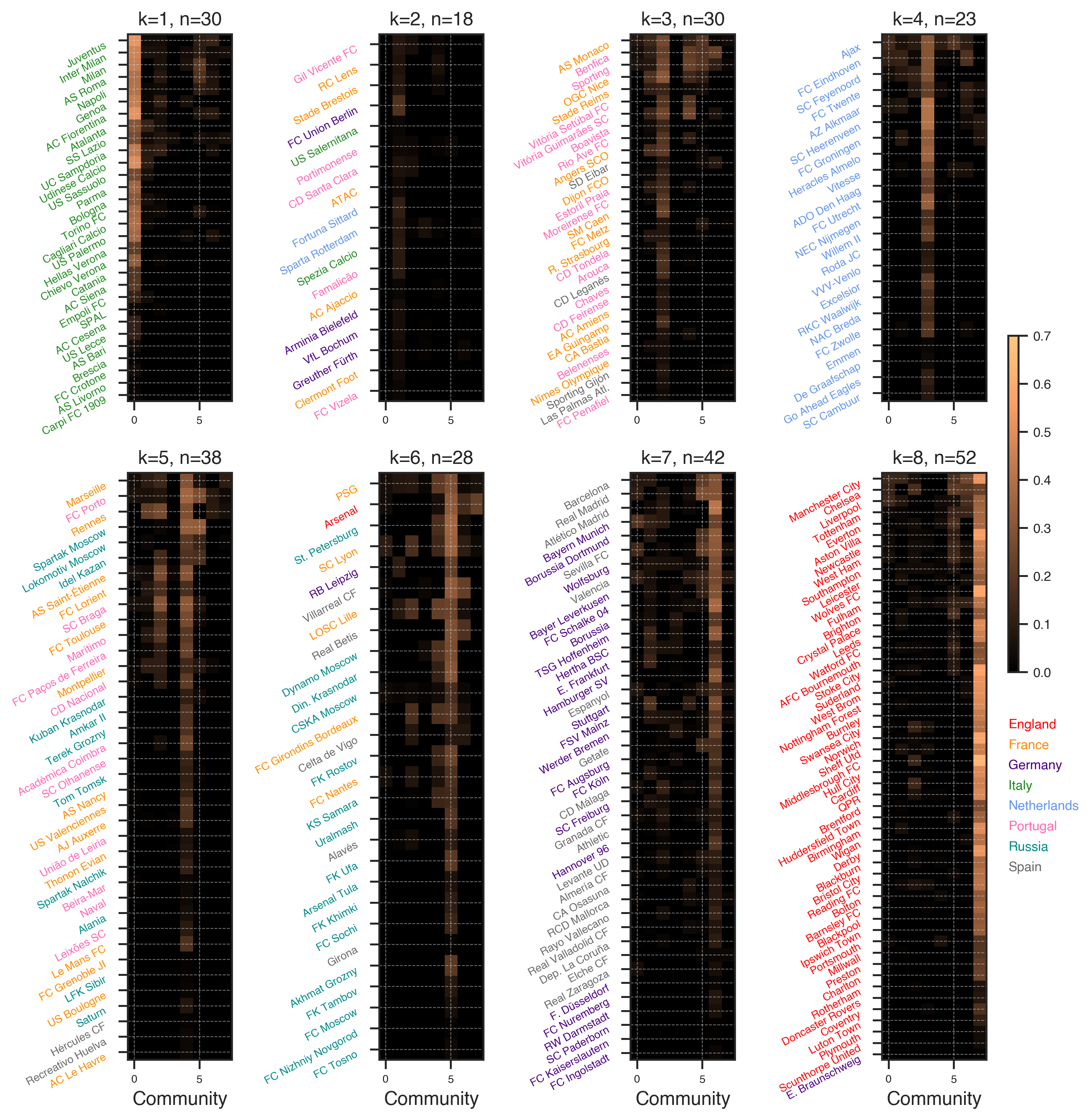}      
\caption{\textbf{Transfermarkt Datasets:} Visualization of the out-going (soft) community membership $u$ of clubs. The colors of the y-labels indicate the country to which the clubs belong. This plot reveals a clear alignment between the community membership of clubs and their respective nationalities. The corresponding country of each league is shown in the legend on the right with the color assigned to that league.\\}
\label{figSI:Transfermarket_community_u}  
\end{figure}

\clearpage
\bibliography{bibliography}
\end{document}

%% file: preamble_math.tex
\newcommand{\be}{\begin{equation}}
\newcommand{\ee}{\end{equation}} 
\newcommand{\bea}{\begin{eqnarray}}
\newcommand{\eea}{\end{eqnarray}}

\newcommand{\e}{\mathrm{e}}

\newcommand{\bern}{\textrm{Bern}}

\newcommand{\pois}{\textrm{Pois}}

\renewcommand{\bf}[1]{\textbf{#1}} % for bold words

\newcommand{\f}[2]{\frac{#1}{#2}}
\newcommand{\ccup}[1]{\left\{#1\right\}}

\newcommand{\rup}[1]{\left[#1\right]}

\newcommand{\Exp}{\mathbb{E}}

\renewcommand{\ref}[1]{[\ref{#1}]}

%% file: algo_dyn_acd.tex
%% ------------------------------ Algorithm ------------------------------------------------------------
\setlength{\textfloatsep}{5pt}
\begin{algorithm}[H]\label{alg:EM}
 \caption{\dmacd:  EM algorithm.}
\SetKwInOut{Input}{Input}
	\setstretch{0.7}
	\Input{network $A(t)=\{A_{ij}(t)\}_{i,j=1}^{N}$, $t=0,\dots,T$ \\number of communities $K$.}
  	\BlankLine
	\KwOut{membership $u=\rup{u_{ik}},\, v=\rup{v_{ik}}$; network affinity matrix $w(t)=\rup{w_{kq}(t)}$; regular edge disappearance rate $\beta(t)$; mean value of Poisson anomaly distribution $\pi$;  prior on anomaly indicator $\mu$; anomalous edge disappearance rate $\phi$, and $\ell$.}
	\BlankLine
	 Initialize $u,v,w(t), \beta, \mu, \pi, \phi,\ell$ at random. 
	 \BlankLine
	 Repeat until $\mathcal{L}$ converges:
	 \BlankLine
	\quad 1. Calculate $\rho$ and $\boldsymbol Q$ (E-step):
	\bea
	 && \rho_{ijkq}(t)=\frac{u_{ik}v_{jq}w_{kq}(t) }{\sum _{k,q}  u_{ik} v_{jq} w_{kq}(t)}  \;,\quad \nonumber \\ &&Q_{ij}  \, \sim \text{as in Eq.} (\ref{appendix:eqn:Qij}) \;.\quad  
	\nonumber
	\eea
%	\be
%	\phi_{ijkq} =\frac{u_{ik}v_{jq}w_{kq}}{\lambda_{ij}^{0}} \nonumber \quad.
%	\ee
	 \quad 2. Update parameters $\Theta$ (M-step):  
	\BlankLine
	\quad \quad \quad 
		i) for each node $i$ and community $k$ update memberships:
		\bea
		\quad  u_{ik}= \f{\sum _{j,q}\, [1-Q_{ij}]\,\sum_{t=0}^T \, \rho_{ijkq}(t)\,\hat{A}_{ij}(t) }{\sum _{j,q}   v _{jq}\, [1-Q_{ij}]\sum_{t=0}^T  \hat{\beta}(t) \,w_{kq}(t)} \quad, \nonumber\\\nonumber\\
\quad  v_{ik} =\f{\sum _{j,q}\,  [1-Q_{ij}]\, \sum_{t=0}^T \, \rho_{ijkq}(t)\,\hat{A}_{ji}(t) }{\sum _{j,q}   u_{jq} \,[1-Q_{ij}]\sum_{t=0}^T  \hat{\beta}(t) \,w_{kq}(t)}\quad,  \nonumber
		\eea
	\quad \quad \quad
	ii) for each pair $(k,q)$ update affinity matrix:
		\be
		\quad w_{kq}(t)=\f{\sum_{i,j}\, [1-Q_{ij}]\,\rho_{ijkq}\,\hat{A}_{ij}(t)}{\sum _{i,j}\, [1-Q_{ij}]\hat{\beta}(t) \, u_{ik}\,v_{jq}} \quad,  \nonumber
		 \ee
	\quad \quad \quad
iii) update  prior on anomaly indicator:
		\be \label{eqn:mu}
		\mu = \f{1}{N(N-1)/2}\sum_{i<j} Q_{ij} \quad,
		\ee
		\quad \quad \quad 
       iv) update anomaly parameters:
       
			\be \label{eqn:pi}
			\pi = \f{\sum_{i,j} \, Q_{ij} \hat{A}_{ij}(0)}{\sum_{i,j}  Q_{ij}}\quad,
			\ee
		\quad \quad \quad 

     \be \label{eqn:ell} 
      \ell = \f{\sum_{i,j}\, \sum_{t=1}^T\,Q_{ij}\, \hat{A}_{ij}(t)}{T\,\sum_{i,j}\, \phi \, Q_{ij}}\quad,
     \ee
     \quad \quad \quad 
     v) The disappearance rate of regular edges $\beta$ has no closed-form update as in Eq. (\ref{appendix:eqn:beta}) \quad,   \\ 
 \quad \quad \quad 	 vi) The disappearance rate of anomalous edges $\phi$ has no closed-form update as in Eq. (\ref{appendix:eqn:phi}) \quad.  \quad \quad \quad
\end{algorithm}
%% ------------------------------------------------------------------------------------------